\documentclass[journal]{IEEEtran}
\usepackage[utf8]{inputenc}
\usepackage{amsmath}

\usepackage{amsfonts}
\usepackage{cite}
\usepackage{dirtytalk}
\usepackage[ruled,vlined,linesnumbered]{algorithm2e}
\usepackage{algorithmic}
\usepackage{xcolor}
\usepackage{colortbl}
\usepackage{lipsum}
\usepackage{bm}
\usepackage{svg}
\DeclareGraphicsExtensions{.png,.pdf}
\usepackage{caption}
\usepackage{subcaption}
\usepackage{multirow}
\usepackage{float}
\usepackage{graphicx}
\graphicspath{{figures/}}
\usepackage{graphicx}
\usepackage{tabularx,booktabs}
\newcolumntype{C}{>{\centering\arraybackslash}X}
\usepackage{makecell}

\usepackage{subfiles} 
\usepackage{multicol}
\usepackage{blindtext}
\usepackage{kantlipsum}
\usepackage{booktabs,threeparttable}
\usepackage{soul}
\usepackage{longtable}

\begin{document}

    \title{A Data-Driven 
    Forced Oscillation Locating Method for Power Systems with Inverter-Based Resources}
    \date{October 2020}
    \author{Yaojie~Cai, Georgia Pierrou,~\IEEEmembership{Member,~IEEE,}
    Xiaozhe Wang,~\IEEEmembership{Senior Member,~IEEE,} and 
    Geza Joos,~\IEEEmembership{Life Fellow,~IEEE} 
     \thanks{ This work was supported in part by the Fonds de recherche du Québec – Nature et technologies (FRQNT) under Grant 256837 and 259083, and in part by the Natural Sciences and Engineering Research Council of Canada (NSERC) under Discovery Grant RGPIN-2022-03236}
    }


    \maketitle
        
    \begin{abstract}

        Forced Oscillations (FO) 
        stemming from external periodic disturbances threaten power system security and stability. The increasing penetration of Inverter-Based Resources (IBRs) further introduces FO, leading to new challenges in identifying and locating FO sources in modern power systems. 
        In this paper,  a novel data-driven method for locating FO in power systems with IBRs is proposed. Unlike previous works, a unified representation of FO originating from IBRs is considered, which further facilitates the development of the FO locating algorithm. Leveraging on Sparse Identification for a Nonlinear Dynamical (SINDy), a purely data-driven methodology is developed for locating the source of FO by interpreting the proposed model from measurements. Numerical results on the WECC 240-bus system validate the performance of the proposed approach in successfully locating FO in the presence of IBRs. \color{black}

        \color{black}
    \end{abstract}

\vspace{-10pt}
\section{Introduction}
Forced Oscillations (FO), \color{black}i.e, oscillations that may appear due to external periodic disturbances, \color{black} pose a significant threat to the security and stability of electric power systems. 
Relevant studies have mostly focused on FO introduced by synchronous generators, for instance, in cases where a 
periodic forcing is imposed on the generator shaft due to inappropriate parameters in the control loop from turbine
governor or exciter.

\color{black}
In recent years, real life events indicate that FO may also originate from Inverter-Based Resources (IBRs). In fact, it has been shown in 
\cite{guideline2017forced} 
that abnormal renewable generator
operating conditions, such as wind shear and solar irradiance
changes, may lead to FO. Besides, when the \color{black}FO \color{black} frequency is low and close to the system modes, it can create resonance conditions with significant oscillations occurring far from the source. 
More such events have been observed with the increasing deployment of IBRs \cite{doubleHeartAttack}, highlighting the need to identify FO sources originating from converter stations.  Consequently, it is of paramount importance that FO are properly identified and their source is located in power systems with both conventional generation and 
 IBRs. 
 


\color{black}

One approach to locating the source of FO is the Dissipating Energy Flow (DEF) method, which identifies the bus with the highest positive DEF 
among generation buses
\color{black} as the FO source \cite{chen2012energy}.  
Although effective in many scenarios, DEF struggles with oscillations from synchronous generator excitation control 
\cite{zhi2020analysis}. 
To address this, the Complex Dissipating Energy Flow (CDEF) method was proposed \cite{estevez2022complex}, introducing a path-dependent energy function that incorporates line conductance and bus voltage changes, significantly improving DEF's accuracy.
In addition, it is tested in \cite{fan2023oscillation} that DEF can be used in the presence of converter-based generation if the complex energy flow is calculated using D-Q axis measurements. 
\color{black}However, both DEF and CDEF methods assume that lines are lossless, which may not be practical. Indeed, it has been shown in \cite{wang2017location} that the DEF method might become less effective 
when calculating the 
DEF through lossy transmission lines.

To locate the source of the FO with minimal assumptions, 
machine learning-based approaches \cite{9211519,9659101,huang2020synchrophasor,cai2022, FENG2022107577} have been widely applied.
K-nearest neighbours classification  \cite{9211519} and deep transfer learning algorithm \cite{FENG2022107577} show promising results in numerical studies. However, the aforementioned methods require a large amount of training data and retraining is needed before deployment to a new system.
Apart from locating the source using blackbox network equations, several methods focus on dynamic components, assuming the system can be represented as low-rank using measurement data. 
The authors in \cite{huang2020synchrophasor} show that FO can be analyzed with a low-rank foreground matrix representing system modes and a sparse background matrix for the corresponding forced input. 
However, when dealing with a large-scale power system, the size of the matrix and consequently the computational demands can rise sharply. 
In \cite{cai2022}, the Sparse Identification of Nonlinear Dynamics (SINDy) methodology introduced 
 by \cite{brunton2016sparse} is utilized to develop a FO locating method in a system with synchronous generators. Yet, 
 locating FO originating from IBRs is not modeled or considered. \color{black}

To study FO in systems with IBRs, \cite{10150043} extends the swing equation that describes synchronous generation dynamics for representing wind power.
Nevertheless, the swing equation cannot capture the dynamics of IBRs, as they do not have rotating masses and only interact with the grid through different electronic controls. 
Thus, a unified model that reflects the convector control characteristics is missing.
Another promising methodology that utilizes the Variational Mode Decomposition (VMD) and Cross-Power Spectral Density (CPSD) 
is presented in \cite{DenisOsipov}. 
Although FO originating from High Voltage Direct Current (HVDC) systems are included, further testing is needed to assess the effectiveness of the VMD-CPSD method in identifying FO from IBRs.\color{black}

\color{black}
To address the aforementioned issues, 
this paper presents a novel data-driven method for locating FO in power systems with IBRs. 
Different from \cite{cai2022} that focuses solely on FO from synchronous generators, this work integrates the presence of IBR devices, such as HVDC and renewable generation, that can also lead to FO\color{black}. In particular, a unified representation of a converter model for FO studies based on Phase-Locked Loop (PLL) dynamics is considered, which facilitates having a common formulation of FO originating from various grid-following (GFL) \color{black} IBRs.
\color{black}
By utilizing SINDy, the sources of FO 
can be identified by extracting the model coefficients from measurements. The effectiveness of the proposed method has been validated through a comprehensive numerical study conducted on the WECC 240-bus system \cite{yuan2020developing}, including scenarios of FO originating from HVDC converter stations and renewable generation converters. \color{black}
In sum, the main contributions of this paper are presented below: 
\begin{enumerate}
    \item A unified representation for FO sourced from GFL \color{black} IBRs is proposed, allowing for the development of a data-driven FO locating algorithm for systems with both synchronous generators and GFL \color{black}IBRs. 
    \item  An extended SINDy function library is designed to effectively integrate GFL \color{black} IBRs as potential FO sources.\color{black}
        \item The proposed method is purely data-driven and requires no additional dynamic model or network parameters.
    \item Comparisons of the proposed approach with respect to state-of-the-art methods are performed on the WECC 240 bus system, which further illustrates the effectiveness of the proposed approach under resonance conditions and non-stationary FO of multiple frequencies.
 \end{enumerate}   
\color{black}The remainder of the paper is organized as follows: Section \ref{section:2} reviews the preliminaries regarding the dynamic model and FO study in the case of synchronous generators,
Section \ref{section:3} describes the FO mechanisms from GFL \color{black} IBRs.
Section \ref{section:4} describes the proposed FO locating method. Section \ref{section:5} presents case studies for validation. Section \ref{section:6} summarizes the conclusions and perspectives for future work.

\color{black}


\color{black}\section{Preliminaries}
\label{section:2}

\subsection{Dynamic Model of Synchronous Generator}

\normalsize

The classical synchronous generator model (i.e., the swing equation) is considered in this study as we are interested in system dynamics in ambient conditions. In particular, the equations which govern generator rotor angles and speed deviations due to stochastic load variations can be expressed as \cite{wang2017pmu}, 
\cite{9115088}: 
        \begin{equation} \label{eq:swingEquation_org}
            \begin{split}
                \bm{\dot{\delta}} &= \bm{\omega} \\
                M\bm{\dot{\omega}} &=   \bm{P_{m}} - \bm{P_{e}} - {D} \bm{\omega} - 
                \Sigma \bm{\eta} \vspace{-2ex} \\                       
            \end{split}
        \end{equation}
\normalsize
\color{black} 
\color{black}
\color{black}where  \color{black}$\bm{\delta}=[\delta_{1}, \cdots, \delta_{\iota}]^T$ is the vector of rotor angles, $\bm{\omega}=[\omega_{1}, \cdots, \omega_{\iota}]^T$ is the vector of rotor speed deviations from the synchronous speed, $\bm{P_{e}}=[P_{e_{1}}, \cdots, P_{e_{\iota}}]^T$ denotes the vector of electrical
power outputs, $\bm{P_{m}} = [P_{m_{1}}, \cdots,  P_{m_{\iota}}]^T$ denotes the vector of mechanical power inputs, and $D=\mbox{diag}[D_1, \cdots, D_\iota]$ is the matrix that includes the damping coefficients. In addition, $\bm{\eta}=[\eta_1,  \cdots,  \eta_\iota ]^T$ is a standard Gaussian random vector and ${\Sigma}=\mbox{diag}[\sigma_{{1}},  \cdots,   \sigma_{{\iota}} ]$ represents the corresponding standard deviations that describe the load power variations reflected on the generation side.

\color{black} 
In FO study, an input $\bm{u_{\omega F}}$ may represent the periodic forcing imposed on the generator turbine shafts and therefore mechanical power $\bm{P_m}$ due to, for example, oscillatory load, mistuned control parameters, or malfunction equipment 
\cite{guideline2017forced}. \color{black}
Thus, the dynamic model of synchronous generators under FO can be represented as: \color{black}

\color{black}

\color{black}
\small
    \begin{equation} \label{eq:swingEqVec2}
    \begin{split}
        \left[\begin{array}{c}
        \bm{\Delta}\bm{ \dot{\delta}} \\
         \bm{\Delta}\bm{ \dot{\omega}}
        \end{array}\right]
        =&
        \left[\begin{array}{cc}
        0_{\iota\times \iota} & I_{\iota\times \iota} \\
        -M^{-1} \frac{\partial \bm{P_{e}}}{\partial \bm{\delta}} &-M^{-1} D
        \end{array}\right]
        \left[\begin{array}{c}
         \bm{\Delta}\bm{ \delta} \\
         \bm{\Delta}\bm{ \omega}
        \end{array}\right] 
        \\
        &-
        \left[\begin{array}{c}
        \bm{0_{\iota\times 1}} \\
        M^{-1}\Sigma \bm{\eta}
        \end{array}\right] 
        +\left[\begin{array}{cc}
        \bm{ u_{\omega F}} 
        \end{array}\right]
    \end{split}
    \end{equation}
\normalsize
where $\Delta$ denotes the difference between the currently measured and the steady state value, and  $\bm{ u_{\omega F}} $ can be represented as: 
\small
     \color{black}
     \begin{equation} \label{eq:inputFO}
     \begin{split}
             \color{black}\bm{u_{\omega F}}\color{black}=\sum_{i=1}^{l}\Bigl(\left[\begin{array}{ccccc}
             \bm{a_{\delta_i}} \\
             \bm{a_{\omega_i}} 
             \end{array}\right] \sin \left(\omega_{F_{i}} t\right)+ 
             \left[\begin{array}{ccccc}
             \bm{b_{\delta_i}} \\
             \bm{b_{\omega_i}} 
             \end{array}\right] \cos \left(\omega_{F_{i}} t\right) \Bigl) 
             \\
            \end{split}
     \end{equation}
     \color{black}
\normalsize
\color{black} 
where \color{black}$\bm{a_{\delta_i}},\bm{a_{\omega_i}},\bm{b_{\delta_i}},\bm{b_{\omega_i}}$ are $\iota \times 1$ vectors that include the coefficients describing the magnitudes of each generator's input at the corresponding frequency $\omega_{F_{i}}, i=1,...,l$.
As discussed in our previous work \cite{cai2022}, FO generated by excitation systems, including exciters, Power System Stabilizers (PSSs), and turbine governors, will manifest themselves in rotor dynamics. Therefore, the model is sufficient to capture FO sourced from synchronous generators.
\color{black}

\subsection{Forced Oscillation Locating Method for Systems with Synchronous Generators}

\color{black}Looking at \eqref{eq:swingEqVec2}, it can be observed that in case the system has a single FO source, the corresponding state with a small amount of trigonometric function coefficients is non-zero. It has been shown in \cite{cai2022} that locating the FO in the context of power systems when the FO source is a synchronous generator can be achieved by levering on measurements and SINDy developed by Brunton et al. in \cite{brunton2016sparse}. Let $\bm{x}=[\bm{\Delta}\bm{ {\delta}}^T, \bm{\Delta}\bm{ {\omega}}^T]^{T}$, and  $X$ contains $\tau_\omega$ snapshots of the state variable $\bm{x}$  and takes the form $X=[\bm{x}_{1},...,\bm{x}_{\tau_\omega}]^T$, \color{black}Briefly, 
the system \eqref{eq:swingEqVec2} can be written in the SINDy form as:
\begin{equation}
\label{eq:sindyform}
    \dot{X}=
    \Xi \Theta^T(X) 
\end{equation}

where the coefficient matrix $\Xi$  and the feature library $\Theta(X)$ are given as: 
\small
    \begin{equation}
        \begin{split}
        \label{eq:Xi}
        \Xi =\left[\begin{array}{lll}
        {\Xi_{\eta}} & \Xi_{\text {Jacobian}} & \Xi_{a b}
        \end{array}\right]\\
        \end{split}
    \end{equation}
    \begin{equation} \label{Xi_DCbias}
        \begin{split}
 {\Xi_{\eta}}=\left[\begin{array}{c}
        0 \\
        -M^{-1} E^{2} G \Sigma \eta \\
        \end{array}\right]
        \end{split}
    \end{equation}
    \begin{equation} \label{Xi_Jacobian}
        \begin{split}
        \Xi_{\text {Jacobian}}=\left[\begin{array}{cc}
        0 & I \\
        -M^{-1} \frac{\partial P_{e}}{\partial \delta} & -M^{-1} D
        \end{array}\right]
        \end{split}
    \end{equation}
    \begin{equation} \label{Xi_ab}
        \begin{split}
        \Xi_{a b}
         =\left[\begin{array}{ccccc}
        |&|&...&|&|\\ 
        {a_{\bm{\delta}_{1}}} & b_{\bm{\delta}_{1}} & \dots &  a_{\bm{\delta}_{l}} & b_{\bm{\delta}_{l}}   \\
        |&|&...&|&|\\ 
        a_{\bm{\omega}_{1}} & b_{\bm{\omega}_{1}} & \dots &  a_{\bm{\omega}_{l}} & b_{\bm{\omega}_{l}}   \\
        |&|&...&|&|\\
        \end{array}\right]
        \end{split}
    \end{equation}
\normalsize

\color{black}

\small
        \begin{equation}
            \Theta^T(X) = 
            \left[\begin{array}{ccc}
                - & \bm{1}& -\\
                - & {X} & - \\
                - & \sin \left(\omega_{F_{1}} t\right) & - \\
                - & \cos \left(\omega_{F_{1}} t\right) & -\\
               \vdots & \vdots& \vdots \\
                - & \sin \left(\omega_{F_{l}} t\right) & -\\
                - & \cos \left(\omega_{F_{l}} t\right)& -
            \end{array}\right]
           \label{eq:systemSINDy}
        \end{equation}
\normalsize
\color{black}
\color{black}Yet, in practice, the success of the source-locating algorithm depends on the sufficient and reasonable feature terms in the SINDy function library. The classic generator model used in \cite{cai2022} cannot accurately represent modern power systems with integrated IBR devices, such as HVDC and renewable generation sources, \color{black} which include GFL \color{black} control and are decoupled from machine dynamics through a DC link. \color{black}
A common dynamic model is needed to allow \cite{cai2022} to further locate the FO source from a GFL \color{black} IBR device with an unknown control loop structure. 
Thus, in the next sections, this paper proposes a general set of dynamic governing equations for locating different FO in power systems,  originating from either conventional synchronous generators or inverter-based renewable generation and the converter control.

\vspace{-10pt}
\color{black}
\section{Forced Oscillation Mechanisms From Inverter-Based Resources}
\label{section:3}

\subsection{Dynamic Model of Grid-Following \color{black} Inverter-Based Resources}

\begin{figure}[!b]
\centering
    \includegraphics[scale = 0.35]{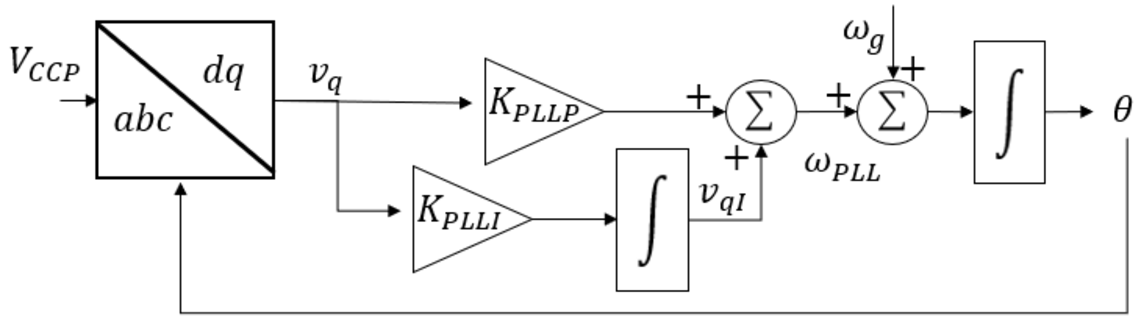}
        \caption{PI Control-based PLL block diagram. }
    \label{fig:selfdraw_PLL2}
\end{figure}

The generalized angle dynamics for GFL \color{black} IBRs \cite{gu2022power} 
demonstrate that the dynamics of the PLL are a major contributor to the overall angle dynamics and are implemented in most grid-following IBRs.
%
A standard synchronous reference frame PLL with a proportional-integral (PI) controller (PI-PLL)  \cite{gu2022power} is illustrated in Fig. \ref{fig:selfdraw_PLL2}. 
In particular, the Q-axis component ${v_q}$ is derived from the three-phase AC voltage at the Common Coupling Point (CCP)  ${V_{ccp}}$ by applying Park's transformation.
A PI controller with proportional and integral gains, $K_{PLLP}$ and $K_{PLLI}$, is employed for low-pass filtering whereas $\omega_g$ represents the system's nominal frequency.

Thus, the PLL equation for the dynamics of the angle $\theta$ can be described as: 
\small
    \begin{equation} \label{eq:pll2}
        \begin{split} 
            \dot{\theta} &=  K_{\text {PLLP }} v_q + K_{\text {PLLI }} \int{v_q }
        \end{split}
    \end{equation}
\normalsize
\color{black}
Assuming the system is connected with $r$ IBR devices, 
(\ref{eq:pll2}) can be linearized and written in a compact form as follows: 
\small
    \begin{equation} \label{eq:pll6}
        \begin{split} 
                \left[\begin{array}{c} \bm{\Delta \dot{\theta}} \end{array}\right]_{r \times 1}
                = & 
                \left[\begin{array}{cc} \bm{K_{PLLP}} \times I & \bm{K_{PLLI}} \times I \end{array}\right]_{r \times 2r}
                \left[\begin{array}{c} \bm{\Delta v_q} \\ 
                \bm{\Delta v_{qI}}\end{array}\right]_{2r \times 1} 
        \end{split}
    \end{equation}
\normalsize
where $\bm{\theta}, \bm{v_q}, \bm{v_{qI}}$ are vectors that include the angles and the corresponding voltage components and $I$ is the identity matrix. Specifically, as illustrated in Fig. \ref{fig:selfdraw_PLL2}, $\bm{v_{qI}}$ is obtained by $\bm{v_q}$ after applying the integral gain.  


\color{black}


Next, various FO mechanisms from GFL \color{black} IBRs documented in previous literature will be discussed, examining their origins, characteristics, and impacts on system dynamics. 
Despite the diversity of IBR-related mechanisms causing FO, we will show that they can consistently be represented as disturbances injected into $V_{ccp}$ as: 
\small
    \begin{equation} \label{eq:eqVCCPFO2}
        \begin{split}   
            V_{ccp_{FO}} = V_{ccp} + u_{F}
        \end{split}
    \end{equation}
\normalsize
\color{black}where $V_{ccp_{FO}}$ denotes the AC voltage under FO and $u_{F}$ is the disturbance injected in $V_{ccp}$ due to FO. \color{black}This unified representation allows for the development of a data-driven FO locating algorithm for power systems with GFL \color{black}IBRs, as will be shown in Section \ref{section:4}. 
\color{black} 

\color{black}
\subsection{
Forced Oscillation from Wind Generation\color{black}}

The periodic fluctuation of wind 
becomes a potential source of FO. 
If the equivalent wind speed $v_{eq_{ws}}$ under wind shear and tower shadow is the source of the injection of FO, 
it can be represented as 
\cite{dolan2006simulation}:
\small
    \begin{equation} \label{eq:eqWindSpeed3}
        \begin{split} 
           v_{eq_{ws}} &= v_{eq}+ v_{F,wind}
        \end{split}
    \end{equation}
\normalsize
where $v_{eq}$ is the hub-height wind speed and $v_{F,wind}$ is the disturbance observed in the wind speed.
The wind speed variation
will be reflected in the power harvested by the turbine $P_{wind_{FO}}$ as follows: 
\small
    \begin{equation} \label{eq:eqWindPower}
        \begin{split} 
           P_{wind_{FO}} &= \frac{1}{2} \rho A_{swept} {v^3_{eq_{ws}}} C_p \\
        \end{split}
    \end{equation}
\normalsize
where $\rho$ is the air density, $A_{swept}$ is the swept area of the turbine blade, and $C_p$ is the power coefficient \cite{5200696}.
Hence, if $v_{eq_{ws}}$ varies periodically, oscillations will be observed in wind generated power. It is worth noting that it has been shown in \cite{Su13} that the frequency range for low-order
oscillations induced from wind turbine control schemes is 0.388 to 0.775 Hz.
\color{black}

\subsection{
Forced Oscillation from Solar Generation\color{black}}
In addition, periodic fluctuations in solar irradiance can cause FO in solar power injection. 
\color{black}
Event 22 in \cite{guideline2017forced} reports a forced oscillation event associated with rapid changes in solar irradiance. Similarly,  \cite{cloud7418531} shows that sudden increases in irradiance due to cloud edge reflection, particularly under weak grid conditions, can interact with inverter control systems and potentially induce oscillations. Furthermore, \cite{ranaweera2014short}  reports significant irradiance variability due to cloud cover in southern Norway and highlights that such fluctuations that occur in seconds to minutes can affect power quality, leading to voltage dips, flicker, and even frequency oscillations. \color{black}
\cite{kankiewicz2010observed, nelson2010effects} analyze power output data from a PV plant, demonstrating that transient clouds whose speed and size influence the duration of shading events result in power oscillations around 4 Hz. \color{black}
Consequently, similar to the case of the wind speed, the varying solar irradiance $g_{eq_{irr}}$ can be assumed as an equivalent irradiance $g_{eq}$ with a periodic signal injection $g_{F, irr}$\cite{surinkaew2020forced}:

\small
    \begin{equation} \label{eq:eqsolar3}
        \begin{split} 
           g_{eq_{irr}} = g_{eq}+ g_{F, irr}
        \end{split}
    \end{equation}
\normalsize
As a result, the generated solar power $P_{solar_{FO}}$ is affected as follows:  
\small
    \begin{equation} \label{eq:eqSolarPower}
        \begin{split} 
           P_{solar_{FO}} &= A_{panel} C_{solar} g_{eq,irr}\\
        \end{split}
    \end{equation}
\normalsize
where $A_{panel}$ is the area and $C_{solar}$ is the efficiency of the solar panel.

\color{black}

\begin{figure}[!t]
\centering
    \includegraphics[width=3.6 in]{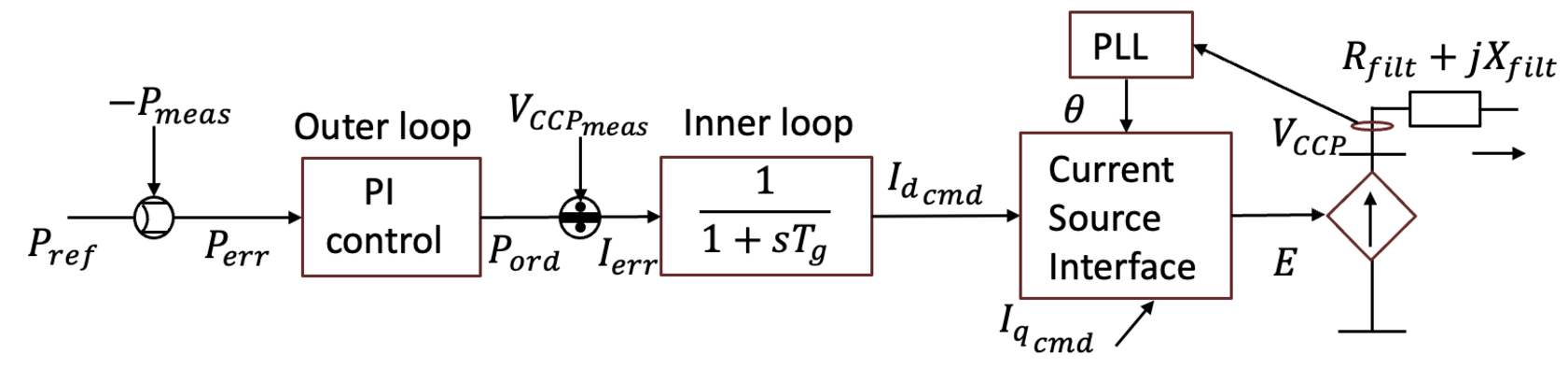}
        \caption{Simplified converter control block diagram. }
    \label{fig:selfdraw_PLL}
    \vspace{-5pt}
\end{figure}

\begin{figure}[!thb]
\vspace{-10pt}
\centering
    \includegraphics[width=3.6 in]{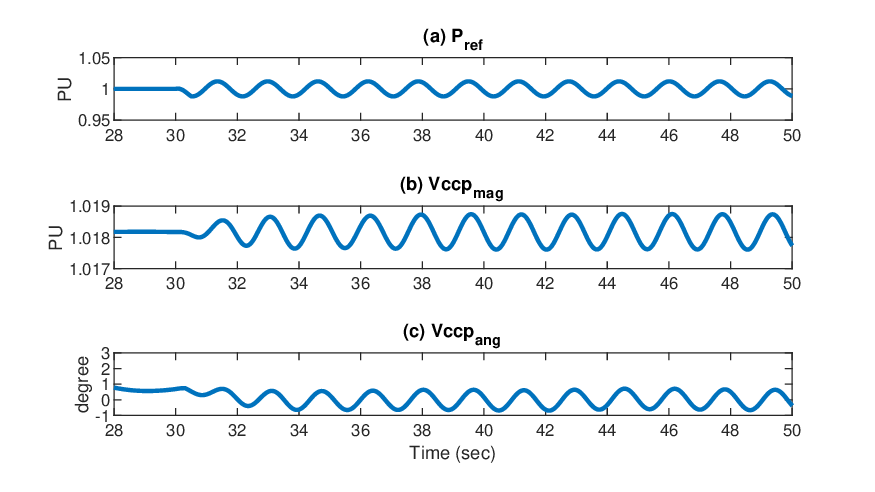}
        \caption{\color{black}(a) Active power reference (b) $V_{ccp}$ magnitude (c) $V_{ccp}$ angle under FO by solar in the WECC 240-bus system. }
    \label{fig:PtoV}
\end{figure}

In sum, (\ref{eq:eqWindPower}) and (\ref{eq:eqSolarPower}) demonstrate that FO originating from either wind generators or variations in solar irradiance will be manifested as oscillations in power injections, i.e., $P_{meas}$ in Fig. \ref{fig:selfdraw_PLL}. Consequently, as indicated by the simplified converter control block diagram in Fig. \ref{fig:selfdraw_PLL}, these oscillations will also appear in the measured voltages at the CCP. To further illustrate that, Fig. \ref{fig:PtoV} depicts simulation results in the WECC 240-bus system \cite{yuan2020developing}, 
where oscillations in both the magnitude and angle of the voltage $V_{ccp}$ are observed once a sinusoidal signal is injected to the active power reference of a solar converter.


\color{black}


\subsection{Forced Oscillation from Converter Control}

\color{black}Except for the inherent uncertain characteristics of the wind speed and solar irradiance, many real-world events indicate that IBR devices can also experience FO due to malfunctioning controllers \cite{guideline2017forced}. 
Indeed, previous works (e.g., \cite{fan2022analysis, vandoorn2013voltage}) 
demonstrate that incorrect parameter settings in various control strategies can lead to undesirable periodic signals 
in $V_{ccp}$ 
when system operating conditions change, such as during power ramping or fault occurrences. 

More specifically, \cite{lu2017evaluation} 
shows that the high PLL gain selection for a converter in a poor grid condition may lead to fast, aggressive responses and eventually oscillatory behaviour in $V_{ccp}$. When the IBR shuts down because of overvoltage, \cite{vandoorn2013voltage} shows that the wrong hysteresis function (determined by a trial-and-error strategy) in the curtailment ON-OFF control can result in the voltage falling below the turn-on value, causing the unit to repeat the stop and restart process and producing an unwanted sawtooth wave at $V_{ccp}$. 
Furthermore, \cite{10579072, 7036119} show that long time delays due to analog-to-digital conversions or communication of the control loop actions may result in oscillations in $V_{ccp}$.
In addition, paper \cite{fan2022analysis} identifies that oscillations of very low frequency (around 0.1 Hz) may appear in $V_{ccp}$ when PV power plants ramp up real power due to communication delays in the volt-var control.  

\begin{figure}[!tb]
\centering
\vspace{-10pt}
    \includegraphics[width=3.6 in]{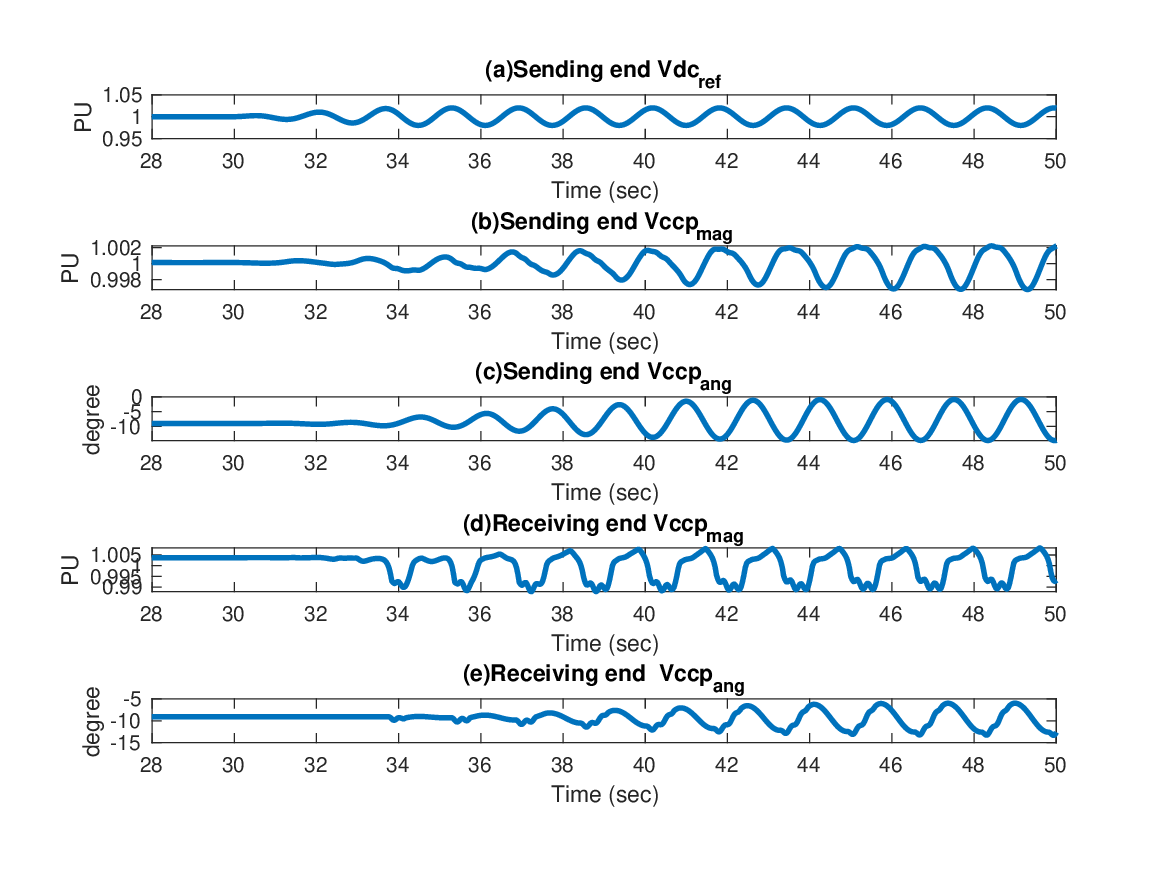}\caption{\color{black}  (a) DC voltage reference, sending end (b and c) and receiving end (d and e) $V_{ccp}$ magnitude and  $V_{ccp}$ angle under FO by HVDC control in the WECC 240-bus system.}
    \label{fig:PtoVDC}
    \vspace{-10pt}
\end{figure}

As further shown in \cite{yuan2020developing}, injecting a sinusoidal signal into the DC voltage control of an HVDC sending end \color{black} converter causes both the magnitude and phase of \color{black} sending and receiving end \color{black}
$V_{ccp}$ \color{black}
to oscillate. To further demonstrate this, Fig. \ref{fig:PtoVDC} presents relevant simulation results obtained from the WECC 240-bus system 
where a FO signal is injected at the DC voltage reference of the sending end bus of the HVDC link. 
Therefore, as proposed by \eqref{eq:eqVCCPFO2}, we will model the oscillations caused by the mistuning of IBR converter control as injected signals into the CCP voltage. Lastly, it is important to note that FO can occur across a broad frequency range. However, this study focuses specifically on low-frequency FO, as these can overlap with the system's natural modes, making source identification particularly challenging. \color{black}
\color{black}

\color{black}
\section{SINDy-based
      Forced Oscillation Locating Method with Grid-Following \color{black} Inverter-Based Resources}
\label{section:4}

\color{black}
\subsection{Dynamic Model of Systems with Grid-Following \color{black}Inverter-Based Resources under Forced Oscillations}


As discussed in the previous section, FO from IBRs will show up at 
$V_{ccp}$. Since 
$v_{q}$ is obtained by applying Park's transformation to $V_{ccp}$, the FO will also appear  
at 
$v_{q}$, which is the input to PLL dynamics as demonstrated by (\ref{eq:pll6}). 
Thus, for $r$ IBR devices, if $\bm{u_F}$ is sourced from 
IBRs, it can be termed as forcing sine and cosine signals added to 
$v_{q}$ and its numerical integration \color{black} $v_{qI}$ in (\ref{eq:pll6}) as below: 

\small
    \begin{equation} \label{eq:pllFO2}
        \begin{split} 
                \left[\begin{array}{c} \bm{\Delta \dot{\theta}} \end{array}\right]
                = & 
                \left[\begin{array}{cc} \bm{K_{PLLP}}\times I  & \bm{K_{PLLI}}\times I  \end{array}\right]_{r \times 2r}  \\ & \times \Bigl(
                \left[\begin{array}{c} \bm{\Delta v_q} \\ 
                \bm{\Delta v_{qI}}\end{array}\right]_{2r \times 1}  
                +                 
                \bm{ u_{F}} \Bigl) \\
                 = &
                \left[\begin{array}{cc} \bm{K_{PLLP}}\times I  & \bm{K_{PLLI}}\times I  \end{array}\right]
                \left[\begin{array}{c} \bm{\Delta v_q} \\ 
                \bm{\Delta v_{qI}}\end{array}\right]  + \\
                &
                \sum_{i=1}^{l}\Bigl(
                    \bm{A_{IBR_i}} 
                 \sin \left(\omega_{F_{i}} t\right)+ \bm{B_{IBR_i}}  \cos \left(\omega_{F_{i}} t\right) \Bigl) 
        \end{split}
    \end{equation}
\normalsize

\color{black}$\bm{A_{{IBR}_i}}$, $\bm{B_{{IBR}_i}}$ are vectors describing each IBR's contribution to each $\omega_{F_{i}}, i=1,...,l$. Specifically,
\color{black}
\begin{equation}
\label{eq:AIBR}
      \bm{A_{IBR_{i}}} = \left[\begin{array}{cc} \bm{K_{PLLP}}\times I  & \bm{K_{PLLI}}\times I  \end{array}\right]                 
      \left[\begin{array}{c} \bm{a_{v_{q}i}}  \\ 
     \bm{a_{v_{qI}i}}\end{array}\right]   
\end{equation}
and
\begin{equation}
\label{eq:BIBR}
     \bm{B_{IBR_{i}}} = \left[\begin{array}{cc} \bm{K_{PLLP}}\times I  & \bm{K_{PLLI}}\times I  \end{array}\right]                 
     \left[\begin{array}{c} \bm{b_{v_{q}i}}  \\ 
     \bm{b_{v_{qI}i}}\end{array}\right]
\end{equation}
where $\bm{a_{v_{q}i}} = [a_{v_{q}i_1},...,a_{v_{q}i_r}]^T$, $\bm{a_{v_{qI}i}} = [a_{v_{qI}i_1},...,a_{v_{qI}i_r}]^T$, $\bm{b_{v_{q}i}} = [b_{v_{q}i_1},...,b_{v_{q}i_r}]^T$, and $\bm{b_{v_{qI}i}} = [b_{v_{qI}i_1},...,b_{v_{qI}i_r}]^T$  are $r \times 1$ vectors that include the coefficients of the IBRs for the corresponding frequency $\omega_{F_{i}}$.

\color{blue} 
\color{black}

\color{black}
\subsection{Forced Oscillation Locating}

The full set of dynamic equations for the bulk power system during FO events incorporates the dynamics of synchronous machines in (\ref{eq:swingEqVec2})  along with the ones of GFL \color{black} IBRs in \eqref{eq:pllFO2}. Thus, the complete model can be expressed in the form of (\ref{eq:sindyform}) as follows, which can be identified using SINDy: \color{black}



\small
    \begin{equation} \label{eq:regressionForm1}
        \begin{split} 
            \begin{aligned}
                \left[\begin{array}{c}
                \bm{\Delta \dot{\delta}} \\
                \bm{\Delta \dot{\theta}} \\
                \bm{\Delta \dot{\omega}}
                \end{array}\right]=\Xi\Theta^T(X)=\Xi\left[\begin{array}{c}
                1 \\
                \bm{\Delta \delta} \\
                \bm{\Delta \omega} \\
                \bm{\Delta v_q} \\
                \bm{\Delta v_{qI}} \\
                \sin \left(\omega_{F_{1}} t\right) \\
                \cos \left(\omega_{F_{1}} t\right) \\
                \vdots \\
                \sin \left(\omega_{F_{l}} t\right) \\
                \cos \left(\omega_{F_{l}} t\right)
                \end{array}\right]
            \end{aligned}
        \end{split}
    \end{equation}
and
\normalsize
\small
        \begin{equation}
            \begin{split}
            \Xi^T =  
                \left[ \begin{array}{cccccc}
                \bm{0} & \bm{0} &  (-M^{-1}  \Sigma \bm{\eta})^T \\   
                \bm{0} & \bm{0} & (-M^{-1} \frac{\partial \bm{P_{e}}}{\partial \bm{\delta} })^T\\
                I &  \bm{0}  & (-M^{-1}D)^T \\
                \bm{0} & \bm{ K_{\text {PLLP }}}\times I & \bm{0} \\
                \bm{0} &  \bm{ K_{\text {PLLI }}}\times I  & \bm{0} \\
                \cellcolor{gray!20}\bm{a_{\delta_{1}}}^T & \cellcolor{gray!20}\bm{A_{IBR_{1}}}^T & \bm{a_{\omega_{1}}}^T\\
                \cellcolor{gray!20}\bm{b_{\delta_{1}}}^T & \cellcolor{gray!20}\bm{B_{IBR_{1}}}^T & \bm{b_{\omega_{1}}}^T\\
                \cellcolor{gray!20}\vdots & \cellcolor{gray!20}\vdots & \vdots\\
                \cellcolor{gray!20}\bm{a_{\delta_{l}}}^T & \cellcolor{gray!20}\bm{A_{IBR_{l}}}^T & \bm{a_{\omega_{l}}}^T\\
                \cellcolor{gray!20}\bm{b_{\delta_{l}}}^T & \cellcolor{gray!20}\bm{B_{IBR_{l}}}^T & \bm{b_{\omega_{l}}}^T
            \end{array}\right]
           \end{split}
           \label{eq:Xi1}
        \end{equation}
\normalsize

\color{black}\noindent where $\bm{a_{\omega_i}}$, $ \bm{b_{\omega_i}}$, $ \bm{a_{\delta_i}}$, $ \bm{b_{\delta_i}}$, $\bm{A_{IBR_i}}$ and $\bm{B_{IBR_i}}$ include the coefficients associated with the sinusoidal terms as introduced in \eqref{eq:inputFO}, \eqref{eq:pllFO2}. Note that  $\bm{\Delta {\delta}}$ and $\bm{\Delta{\omega}}$ are the deviations of the rotor angles and rotor speeds of synchronous generators, respectively, and $\bm{\Delta{\theta}}$ are the deviations of IBRs PLL angles. \color{black}

\color{black}

Recall the SINDy form from Section \ref{section:2},
the $X$ matrix on the left-hand side of (\ref{eq:regressionForm1}) can be built by using either directly obtained or calculated from PMU measurements associated with the state variables of synchronous generators and GFL IBR devices \color{black} with a window size $\tau_{w}$. Let $\bm{x}=[\bm{\Delta}\bm{ {\delta}}^T, \bm{\Delta}\bm{ {\theta}}^T, \bm{\Delta}\bm{ {\omega}}^T]^{T}$, and  $X$ contains $\tau_\omega$ snapshots of the state variable $\bm{x}$:  
    \begin{equation} \label{eq:derivative matrix}
        \begin{split}
            X = 
            \left[
            \begin{array}{ccccccccccccccc}
             |  & |  & \dots & |  \\
            \bm{x}_1 & \bm{x}_2  & \dots & \bm{x}_{\tau_{\omega}} \\
            |  & |  & \dots & |  \\ 
            \end{array}\right]
        \end{split}
    \end{equation}
\color{black}In addition, the right-hand side library matrix $\Theta(X)$ of (\ref{eq:regressionForm1}) is built as:  

    \begin{equation} \label{eq:library}
        \begin{split}
            \Theta(X) = \left[
            \begin{array}{ccccccccccccccc}
            | & \dots & |  & \dots & |  & \dots & |  & \dots\\ 
            \bm{1}  &  \dots & {\Delta \delta}_{\iota} & \dots  & {\Delta \omega}_{\iota} & \dots & {\Delta v}_{q_{r}} & \dots  \\
            | & \dots & |  & \dots & |  & \dots & |  & \dots  \\
            \end{array}\right.
        \end{split}
    \end{equation}
        $\qquad 
        \left.\begin{array}{ccccc}
        \mid & \ldots & \mid & \ldots & \mid \\
        \Delta v_{{qI}_{r}} & \ldots & \sin (\omega_{F_{i}}t) & \ldots & \cos (\omega_{F_{i}}t) \\
        \mid & \ldots & \mid & \ldots & \mid
        \end{array}\right] $ \\ 
Each column in $\Theta(X)$ means $\tau_\omega$ snapshots of the variable. \color{black}The data for the integral control $\bm{v_{qI}}$ can be calculated discretely with the time step, a data buffer record, and numerical integration using the trapezoid or Simpson's rule (\ref{eq:pll5}) \cite{lindfield2018numerical} as:
\small
    \begin{equation} \label{eq:pll5}
        \begin{split} 
            \bm{v_{qI}} =  \int_{t-\tau}^{t} \bm{v_q} \,dt =  \frac{\tau}{2} \left[\bm{v_q}(t) + \bm{v_q}(t-\tau) \right]  
        \end{split}
    \end{equation}
\normalsize
\color{black}
where $\tau$ is the time step of the measurements. 

\color{black}
Next, the Fast Fourier Transform (FFT) is applied to the magnitude and phase angle of each element of $\bm{V_{ccp}}$. By rescaling the maximum amplitude of the spectrum from FFT between $0$ and $1$, the FO frequencies can be obtained through a z-score-based peak detection method \cite{zscore} with a threshold of $1$ from the resulting spectrum for each measurement. 
Up to three ($l\leq3$) potential frequencies are selected from the spectrum, assuming that the injected FO frequencies are among the highest dominant oscillation modes.


\color{black}
Given the time derivative matrix (\ref{eq:derivative matrix}) and the library matrix in (\ref{eq:library}), 
(\ref{eq:Xi1}) can be obtained by a recursive threshold least-squares solution. SINDy promotes sparsity in $\boldsymbol{\Xi}$ as in (\ref{eq:sindyform}) using an iterative Sequential Thresholded Least Squares (STLS) algorithm \cite{brunton2019data}. This approach alternates between least-squares regression and hard thresholding of small coefficients. For each state variable $k = 1, \dots, n$, the update procedure is given by (\ref{eq:stls}) \cite{brunton2019data}:
\begin{equation}
\begin{aligned}
{\xi_k^{(0)}}^T &= \arg\min_{\xi^T} \left\| \dot{X}_k^T - \Theta(X)\xi_k^T \right\|_2^2, \\
\xi_{k,i}^{(j)} &= 0 \quad \text{if} \quad \left| \xi_{k,i}^{(j)} \right| < \lambda, \\
A_k^{(j)} &= \left\{ i : \xi_{k,i}^{(j)} \neq 0 \right\}, \\
\xi_k^{(j+1)^T} &= \arg\min_{\xi^T \in \mathbb{R}^{|A_k^{(j)}|}} \left\| \dot{X}_k^T  - \Theta(X)_{A_k^{(j)}}\xi_k^T \right\|_2^2,
\end{aligned}
\label{eq:stls}
\end{equation}
where 
$\xi_k$ is the $k$-th row of $\Xi$, 
$\xi_{{k,i}}^{(j)}$ is 
the entry in row $k$, column $i$ of $\Xi$ at iteration $j$.
$\lambda > 0$ is a sparsity threshold, and $\Theta(X)_{A_k^{(j)}}$ denotes the submatrix of active terms. This procedure is repeated until the active set $A_k^{(j)}$ stabilizes, i.e., $A_k^{(j+1)} = A_k^{(j)}$.
\color{black}

\color{black}
Looking at \eqref{eq:Xi1}, after obtaining the estimated $\Xi$, the elements from its bottom left region (the shaded area in (\ref{eq:Xi1})), which represent the FO coefficient of angle dynamics from synchronous generators and GFL IBRs can be extracted: 
    \begin{equation}
        \begin{split}
        \xi_{angle} =  
            \left[ \begin{array}{cccccc}
                \bm{a_{\delta_{1}}} & \bm{b_{\delta_{1}}} & \dots & \bm{a_{\delta_{l}}} &  \bm{b_{\delta_{l}}} \\
                \bm{A_{IBR_{1}}} & \bm{B_{IBR_{1}}}   & \dots    & \bm{A_{IBR_{l}}}   & \bm{B_{IBR_{l}}}    
            \end{array}\right]        
       \end{split}
       \label{eq:Xi_ang1}
    \end{equation}

Let $\zeta_{angle}$ be the square summation of the corresponding sinusoidal coefficients of the rotor angles $\bm{\delta}$ and PLL output angles $\bm{\theta}$ in $\xi_{angle}$ as:
    \begin{equation}
        \begin{split}
        \zeta_{angle} =  
            \left[ \begin{array}{cccccc}
                \bm{a_{\delta_{1}}}^{2} + \bm{b_{\delta_{1}}}^{2} & \bm{A_{IBR_{1}}}^{2} + \bm{B_{IBR_{1}}}^{2} \\
                \vdots & \vdots
                \\
                \bm{a_{\delta_{l}}}^{2} + \bm{b_{\delta_{l}}}^{2} & \bm{A_{IBR_{l}}}^{2} + \bm{B_{IBR_{l}}}^{2}      
            \end{array}\right]        
       \end{split}
       \label{eq:Xi_ang2}
    \end{equation}   
Then, each $(i,j)$ element in $\zeta_{angle}$ is the estimated squared magnitude of the injected FO of frequency $\omega_{F_i}$ from the generator or IBR $j$.
Therefore, the FO frequencies $\omega_{F_{i}}$ and the source generators $j$ can be identified from the highest peak values in $\zeta_{angle}$. 
The proposed complete procedure for extracting the FO location is presented in Algorithm 1.
        \small
                \SetKwInput{stepOne}{Step 1}
                \SetKwInput{stepTwo}{Step 2}
                \SetKwInput{stepThree}{Step 3}
                \SetKwInput{stepFour}{Step 4}
                \SetKwInput{stepFive}{Step 5}
                \SetKwInput{stepSex}{Step 6}
                \begin{algorithm}
                    \stepOne{Collect $\bm{\Delta \dot{\delta}}$, $\bm{\Delta \dot{\theta}}$, $\bm{\Delta \dot{\omega}}$, and $\bm{v_{q}}$ measurements \color{black} within a window $\tau_{w}$. 
                     Compute the integral signal by \color{black}(\ref{eq:pll5}). 
                    }
                    \stepTwo{Conduct an FFT analysis on measurements from each $V_{ccp}$. 
                    Apply the z-score-based peak detection techniques to identify the FO frequencies $\bm{\omega_{F}}$. 
                    \color{black}
                    }
                    \stepThree{Build the time derivative matrix $X$ (\ref{eq:derivative matrix}) and the library matrix $\Theta(X)$ (\ref{eq:library}). 
                    }                   
                    \stepFour{Obtain the coefficient matrix $\Xi$ by (\ref{eq:stls}).}
                    \stepFive{Calculate $\zeta_{angle}$ by (\ref{eq:Xi_ang2}) and identify the sources by selecting the element with the highest peak values in  $\zeta_{angle}$.}
                    \caption{SINDy based data-driven method to locate FO in systems with 
                    IBRs}
                \end{algorithm}
        \normalsize

\textbf{Remarks}:\\
\noindent$\bullet$ In \textbf{Step 1}, the method requires rotor angles and rotor speeds of synchronous generators, along with each IBR’s PLL angle and Q-axis voltage component. Notably, if a bus is connected to only a single generator, PMU measurements at the CCP can be used to approximate all required values. However, when multiple generators (either synchronous generators or IBRs) are connected to the same bus, internal generator measurements are preferred for accurately locating the FO source.
\color{black}

\noindent$\bullet$ 
A sufficient number of measurements 
is necessary to be collected at \textbf{Step 1} of the proposed algorithm. In this paper, $3600$ measurements are used, as the window size $\tau_{w}$ 
is $40$ seconds and the sampling time 
is $0.017$ seconds.
The applied window size ensures that FO with low frequencies can be captured accurately.
The two-points forward difference approximation method  \cite{lindfield2018numerical} is applied to estimate the derivatives of measurements. 


\noindent$\bullet$ 
The threshold in \eqref{eq:stls} can be adjusted with respect to the number of the non-zero elements in $\zeta_{angle}$. In this work, the threshold is selected as $0.006$, allowing for three non-zero elements.

\noindent$\bullet$ To calculate the derivative of \eqref{eq:derivative matrix}, in our case, a single differentiation does not significantly degrade numerical stability with the given sampling rate. However, in practice, engineers rarely compute derivatives directly due to noise issues of numerical differentiation, especially for higher-order derivatives or with noisy data. In practice, engineers usually use various noise mitigation strategies \cite{press2007numerical} such as smoothing before derivetive (moving average, low-pass filtering), different finite-difference approximations (central, forward, backward difference), polynomial fitting (Savitzky-Golay filter, spline differentiation, etc.)
, and state estimation techniques (Kalman filtering, total variation regularization, etc.) \cite{press2007numerical}.

\noindent$\bullet$  It is worth noting that the proposed method can inherently locate FO with multiple frequencies 
as all identified frequencies are incorporated into the feature library $\Theta(X)$. The FO sources then manifest as nonzero entries in the gray area of $\Xi$ in \eqref{eq:Xi1}, which SINDy utilizes for source identification. As will be elaborated in Section V-D, this makes the proposed approach advantageous with respect to existing FO locating methods that handle FO from one frequency, e.g., the VMD-CPSD method \cite{DenisOsipov} and the DEF method \cite{maslennikov2020iso}. 

\color{black}

\section{Numerical Results}
\label{section:5}
\begin{figure}[h]
    \centering
    \includegraphics[width=3 in]{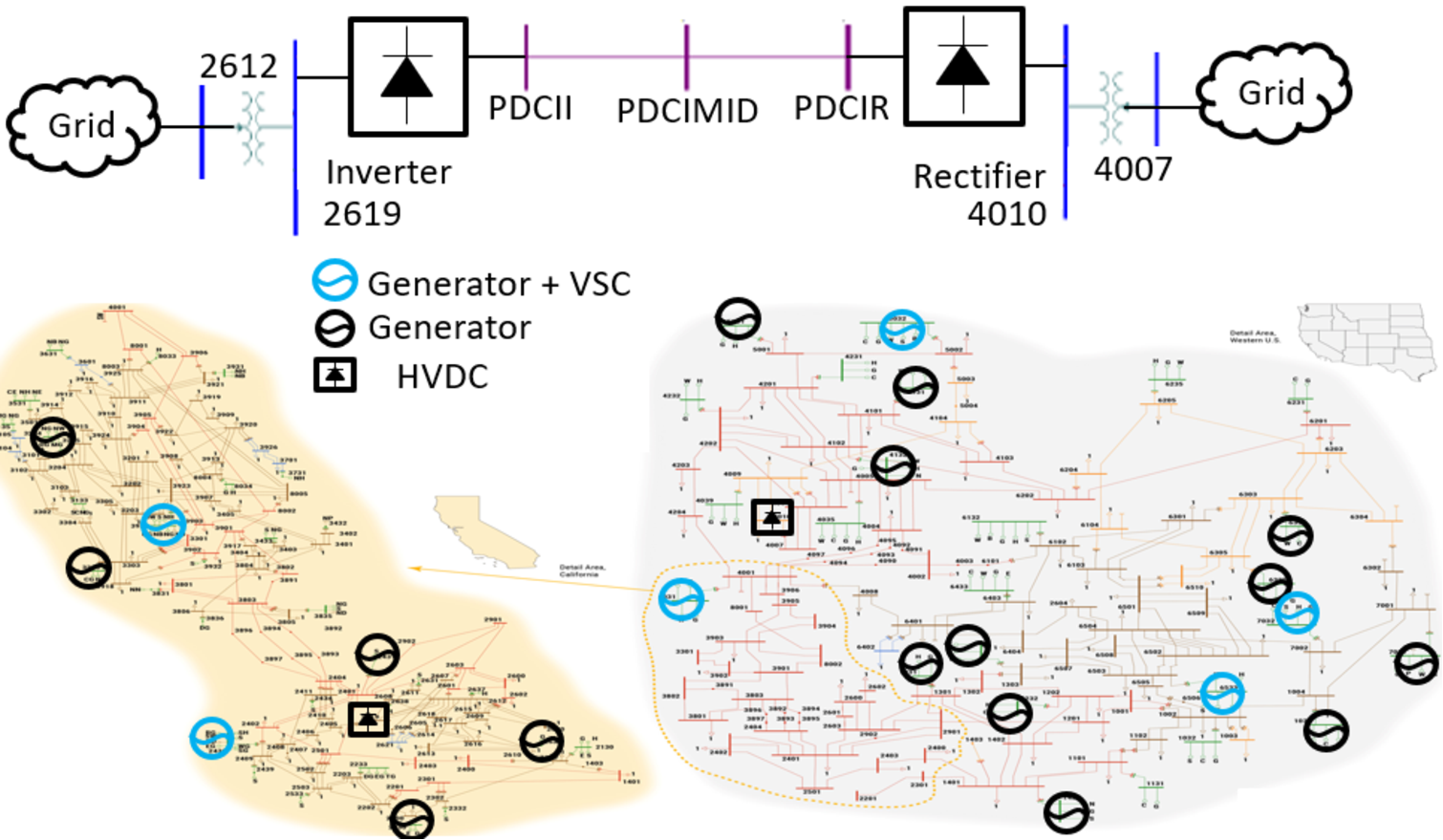}
        \caption{Voltage Source Based  Interface model \cite{tsatPLL}.}
    \label{fig:selfdraw_TSAT_VSC}
\end{figure}

To test the proposed algorithm in an extensive system  \color{black} that includes both GFL \color{black} IBRs and synchronous generators\color{black}, a set of simulations is carried out on the WECC 240-bus test system developed by National Renewable Energy Laboratory (NREL) 
\cite{yuan2020developing} using the commercially available TSAT software \cite{tsatPLL}. 
All synchronous generators are modeled with the 
sixth-order \color{black} machine model ``GENROU" equipped with exciters and turbines 
\color{black}as presented in \cite{tsatPLL}. \color{black}The generator connected to bus $1032$ has been selected as the reference generator.
\color{black}
To integrate IBRs, a Voltage Source Converter (VSC)-HVDC 
connection between buses $4010$ and $2619$   
is considered which consists of the current limiter, PLL, internal voltage calculation, and outer loop control. \color{black} 
\color{black}
Following Algorithm 1, measurements from both synchronous generators and GFL \color{black} IBRs are used. Specifically, $24$ buses with synchronous generators are selected as the monitored buses for generator rotor angle and speed measurements. The outputs of the PLL, $\theta$ and the Q-axis voltage $v_q$, are measured and $v_{qI}$ is calculated by the trapezoid rule (\ref{eq:pll5}). \color{black} 
\color{black}
\color{black} In addition, there are $9$ renewable generation sources including solar generation at Buses $2438$, $3933$, $4031$, $5032$, $6533$, $7032$ and wind generation at Buses $4031$, $5032$, $6533$. \color{black} 
\color{black}
\color{black} An illustration of the monitored system is shown in Fig. \ref{fig:selfdraw_TSAT_VSC}.\color{black}

\subsection{Methods for comparison}
\color{black}
\color{black}
The proposed method is validated and compared to other methods. In particular,  the VMD-CPSD method presented in \cite{DenisOsipov} and the DEF algorithm developed in \cite{maslennikov2020iso} for FO locating are also implemented in TSAT for comparison. 


Briefly, the VMD-CPSD method \cite{DenisOsipov} exploits a cross-power spectral density analysis, focusing on the relationships between voltage magnitude and real/reactive power (SVmP / SVmQ), as well as voltage angle and real power (SVaP). 
It pinpoints the FO source location based on the largest positive imaginary part of the cross-power spectral density. The DEF method \cite{maslennikov2020iso} leverages the concept of energy dissipation within the system. Considering that in a stable power system oscillation energy should dissipate over time, DEF identifies the source of a FO by focusing on the buses where an energy amplification is observed, implying the existence of external forcing. 


\color{black}
\color{black}

\subsection{
Locating \color{black} Forced Oscillations from Renewable Generations}

\color{black}To simulate FO originating from renewable generation, a forced signal is applied to the active power control $P_{ref}$ of various renewable converters at frequencies that coincide with the system's inter-area modes, \color{black}specifically \color{black} 0.379 Hz, 0.614 Hz, and 1.27 Hz, thereby creating resonance conditions.
A total of 27 tests were conducted by injecting FO into each monitored renewable generator at different frequencies. Tables \ref{tab:case_renewable}, \ref{tab:case_renewable2}, and \ref{tab:case_renewable3} present the results for all cases. \color{black} It can be observed that the proposed method successfully locates the FO source in all cases. However, the DEF and VMD-CPSD may fail to properly identify the FO source. Three such cases are highlighted in grey colors, where the DEF and VMD-CPSD methods either identify a nearby bus to the true source (light grey) or a more distant bus (dark grey). Hence, the proposed method outperforms DEF and VMD-CPSD in properly locating the FO sources from wind or solar generation. \color{black} 


\small
\begin{table}[!htbp]
    \caption{\color{black}Forced oscillation cases at $0.379$ 
 Hz originating from solar/wind generation}
    \label{tab:case_renewable}
    \centering
    \begin{tabular}{c c c c c c}
    \toprule
    \multicolumn{3}{|c|}{Simulation setup} & \multicolumn{1}{c|}{DEF}& \multicolumn{1}{c|}{SINDy}& \multicolumn{1}{c|}{VMD-CPSD} \\ 
    \midrule
        {\#}& FO Freq &  Injection bus  &  Est bus  & Est bus  &  Est bus \\ \midrule
        {1} &{} & 6533 solar   & 6533   &  6533   & 6533  \\ 
        {2} &{} &  7032 solar   & 7032   &  7032   & 7032  \\ 
        {3} &{} &  5032 solar     & 5032   &  5032   & 5032  \\ 
        {4} &{0.379Hz} &  4031 solar   & 4031   &  4031   & 4031  \\ 
        {5} &{} &  3933 solar      &  3933   & 3933  &3933\\ 
        {6} &{} &  2438 solar   &  2438  &  2438   & 2438  \\ 
        {7} &{} &  4031 wind   & 4031   &  4031   & 4031  \\ 
        {8} &{} &  5032 wind    & \cellcolor{gray!40}4031   &  5032 &   5032  \\ 
        {9} &{} &  6533 wind    & 6533   &  6533   & 6533  \\     
        \bottomrule
    \end{tabular}
\end{table}
\normalsize

\small
\vspace{5 pt}
\begin{table} [!htbp]
    \caption{\color{black}Forced oscillation cases at $0.614$ Hz originating from solar/wind generation\color{black}}
    \label{tab:case_renewable2}
    \centering
    \begin{tabular}{c c c c c c c}
    \toprule
    \multicolumn{3}{|c|}{Simulation setup} & \multicolumn{1}{c|}{DEF}& \multicolumn{1}{c|}{SINDy}& \multicolumn{1}{c|}{VMD-CPSD} \\ 
    \midrule
        {\#} &FO Freq &  Injection bus  &  Est bus  & Est bus  &  Est bus \\ \midrule
        {1} &{} &  6533 solar   & 6533   &  6533   & \cellcolor{gray!20}6333  \\ 
        {2} &{} &  7032 solar     & 7032   &  7032   &  7032  \\ 
        {3} &{} &  5032 solar     & 5032   &  5032   & 5032  \\ 
        {4} &{} &  4031 solar   & 4031   &  4031   & 4031  \\ 
        {5} &{0.614Hz} & 3933 solar    &  3933   & 3933  & 3933  \\ 
        {6} &{} &  2438 solar   & \cellcolor{gray!20}2619   &  2438   & 2438  \\ 
        {7} &{} &  4031 wind   & 4031   &  4031   & 4031  \\ 
        {8} &{} &  5032 wind   & 5032   &  5032   & 5032  \\ 
        {9} &{} &  6533 wind   & 6533   &  6533   & 6533  \\      
        \bottomrule
    \end{tabular}
\end{table}
\normalsize

\small
\vspace{5 pt}
\begin{table}[!htbp]
    \caption{\color{black}Forced oscillation cases at $1.27$ Hz originating from solar/wind generation}
    \label{tab:case_renewable3}
    \centering
    \begin{tabular}{c c c c c c c}
    \toprule
    \multicolumn{3}{|c|}{Simulation setup} & \multicolumn{1}{c|}{DEF}& \multicolumn{1}{c|}{SINDy}& \multicolumn{1}{c|}{VMD-CPSD} \\ 
    \midrule
        {\#} &FO Freq &  Injection bus  &  Est bus  & Est bus  &  Est bus \\ \midrule
        {1} &{} &  6533 solar    & 6533   &  6533   & 6533  \\ 
        {2} &{} &  7032 solar   & 7032   &  7032   & 7031  \\ 
        {3} &{} & 5032 solar    & 5032   &  5032   & 5032  \\ 
        {4} &{} & 4031 solar   & 4031   &  4031   & 4031  \\ 
        {5} &{1.27Hz} & 3933 solar      &  3933   & 3933 & 3933 \\ 
        {6} &{} &2438 solar   & 2438   &  2438   & 2438  \\ 
        {7} &{} &  4031 wind  & 4031   &  4031   & 4031  \\ 
        {8} &{} &  5032 wind    & 5032   &  5032   & 5032  \\ 
        {9} &{} & 6533 wind   & 6533   &  6533   & 6533  \\   
        \bottomrule
    \end{tabular}
\end{table}
\normalsize
\color{black}



\begin{figure}[!b]
    \centering
    \includegraphics[width=3.2 in]{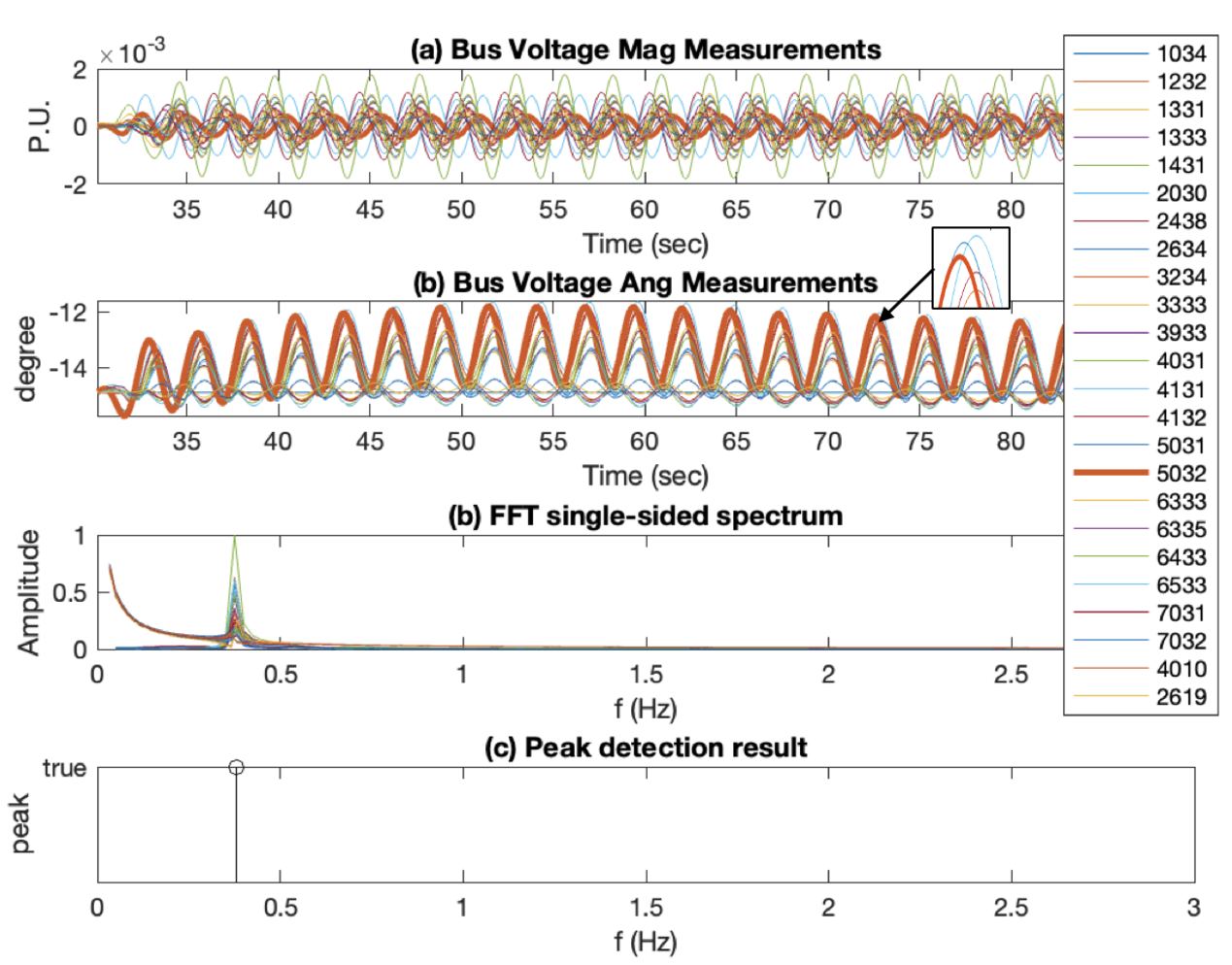}
        \caption{\color{black} (a) $V_{ccp}$ magnitude (b) $V_{ccp}$ angle (c) The single-sided spectrum from FFT (d) Peak frequency detection in the case of forced oscillations originating from wind generation at bus $5032$.}
    \label{fig:StatFO_5032w_ang}
\end{figure}
\begin{figure}[!b]
    \centering
        \includegraphics[width=3.2 in]{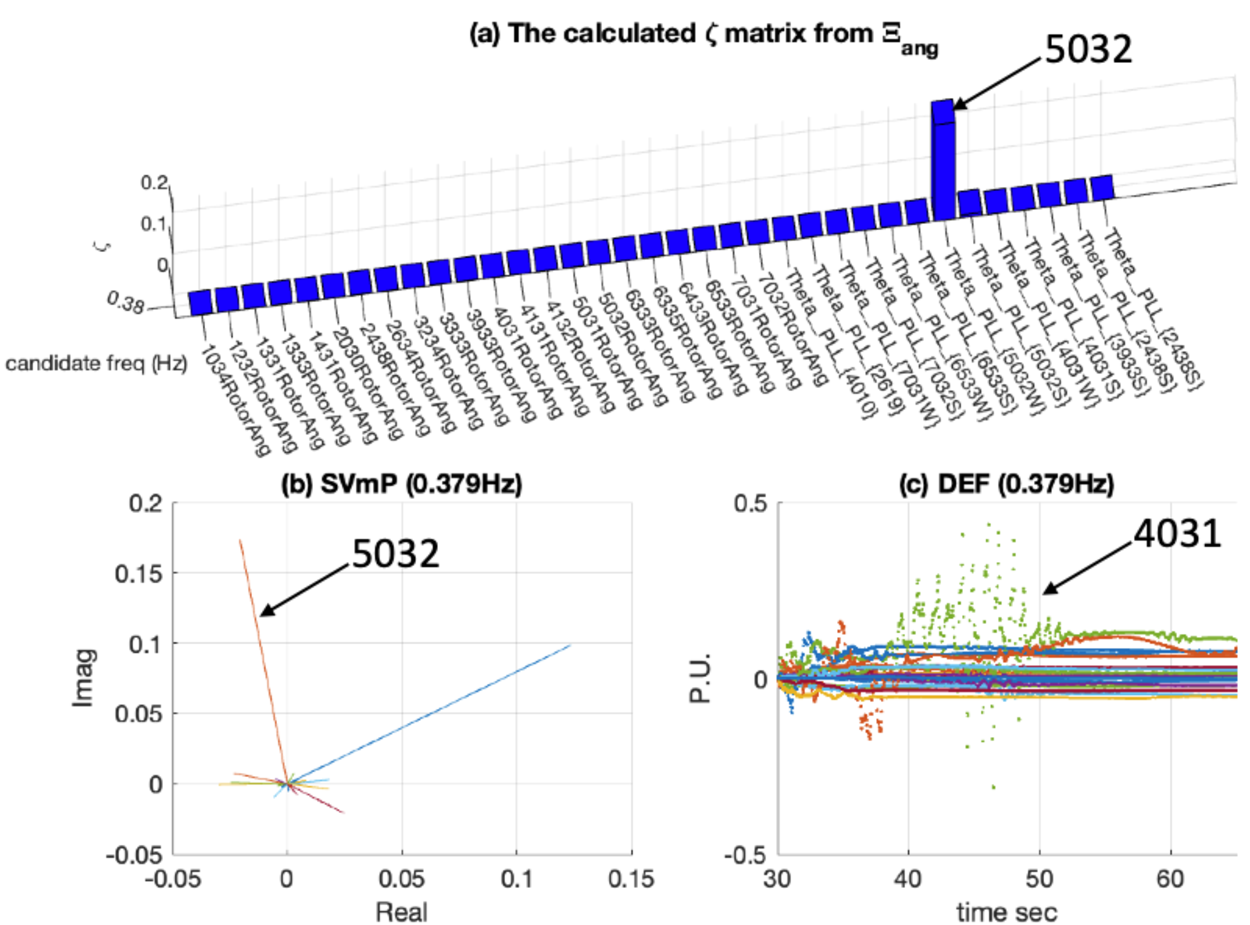}
        \caption{\color{black} (a) Proposed method (b) VMD-CPSD method (c) DEF method in the case of forced oscillations originating from wind generation at bus $5032$. 
        }
    \label{fig:StatFO_5032w_Result}
\end{figure}

\color{black}
To further illustrate this result, we first consider Case 8 in Table \ref{tab:case_renewable}, where the FO is injected to the active power control reference of the wind generation at Bus $5032$. The $V_{ccp}$ magnitude and angle trajectories 
are shown in Fig. \ref{fig:StatFO_5032w_ang} (a) and (b). In addition, the single-sided spectrum from FFT and the associated peak detection observation \cite{zscore} are presented in Fig.\ref{fig:StatFO_5032w_ang} (c) and (d).  
 Fig. \ref{fig:StatFO_5032w_ang} indicates that the oscillation strength is $2.22\%$ of the reference value and the frequency is $0.379$ Hz. 
It is worth noting that despite the fact that the maximum oscillation amplitude is not at the source, the proposed algorithm can properly find the correct source location. As shown in Fig. \ref{fig:StatFO_5032w_Result} (a), this can be achieved by utilizing the rotor angle, speed, and PLL signal from the buses where renewable generation and the HVDC system are located.

Fig. \ref{fig:StatFO_5032w_Result} (b) shows that the VMD-CPSD method successfully identifies the source based on the maximum imaginary parts in SVmP quantities at the FO frequency.
However, as illustrated in Fig. \ref{fig:StatFO_5032w_Result} (c), the source of the FO is misidentified by the DEF method, as the dissipate energy for bus $4031$ increases whereas the FO is injected at bus $5032$. 

\color{black}

\begin{figure}[!t]
    \centering
    \includegraphics[width=3.2 in]{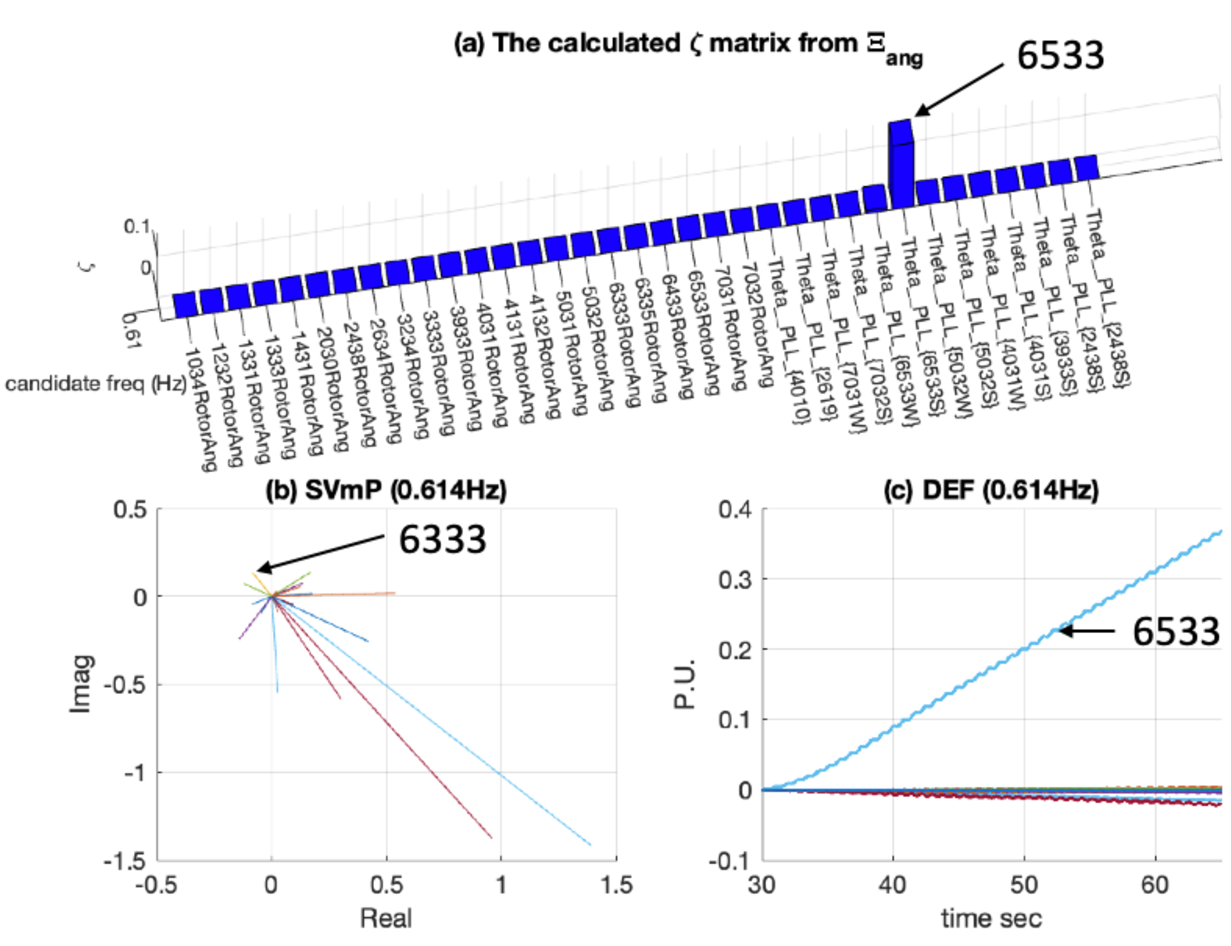}
        \caption{\color{black}(a) Proposed method (b) VMD-CPSD method (c) DEF method in the case of forced oscillations originating from  solar generation at bus $6533$.}
    \label{fig:StatFO_6533s_Result}
\end{figure}

\color{black}
Then, we focus on Case $1$ in Table \ref{tab:case_renewable2}, which corresponds to injecting FO at the outer loop active power control of the solar generation at bus $6533$. From Fig. \ref{fig:StatFO_6533s_Result}, it can be observed that the proposed method and the DEF method can identify the true location of the FO source. Yet, the 
VMD-CPSD method fails to do so since the corresponding metric falsely indicates the nearby bus $6333$ as the FO source. 

\subsection{Forced Oscillations due to Mistuned Controller Parameters}
To test the performance of the proposed method in the case of FO due to mistuned parameters in the IBR-related controllers, 
FO of $0.614$ Hz  are injected to the VSC-HVDC converter models, specifically at the sending end bus $4010$ and the receiving end bus $2619$. 
It should be highlighted that HVDC is used instead of renewable generators because TSAT provides access to their inner loop control through a user-defined HVDC model, whereas the inner loop control of renewable generators is not accessible. 
As summarized in Table \ref{tab:case_HVDC}, the VMD-CPSD method is the most effective in identifying the source of FO related to HVDC converter control. While the proposed SINDy method can determine that the HVDC system is the source, it may not accurately distinguish whether the source is the sending or receiving end. In contrast, DEF often identifies nearby buses around the HVDC converter as the FO sources rather than pinpointing the actual location. \color{blue}
\begin{table*}[!ht]
     \caption{Forced oscillation cases originating from HVDC system \color{black}}
    \label{tab:case_HVDC}
    \centering
    \begin{tabularx}{\textwidth}{c c c c c }
        \toprule
        \multicolumn{2}{|@{}c|}{  \qquad \qquad Source from  HVDC system \qquad \qquad } & \multicolumn{1}{@{}c|}{\qquad \qquad \qquad  DEF Result  \qquad \qquad \qquad }  & \multicolumn{1}{@{}c|}{\qquad \qquad SINDy Result \qquad \qquad} & \multicolumn{1}{@{}c|}{\qquad \qquad  VMD-CPSD Result \qquad \qquad}\\ 
        \midrule
        {\#}  &  Location  &  Est bus  & Est bus  &  Est bus \\ \midrule
        1 &  2619 outer loop real power control  & 2619   &  2619   & 2619  \\ 
        2 &  2619 inner loop Q-axis current reference    & \cellcolor{gray!20}2438   &  2619   & 2619 \\
        3 &  2619 inner loop D-axis current reference    & 2619   &  2619   & 2619  \\ 
        4 & 2619  PLL output angle \color{black}  &  2619   & \cellcolor{gray!20}4010 & \cellcolor{gray!20}4010 \\ 
        5 &  4010  outer loop DC voltage control   &  \cellcolor{gray!20}4131   & 4010 & 4010  \\ 
        6 &  4010 outer loop AC voltage control   & \cellcolor{gray!20}2619  & \cellcolor{gray!20}2619 & 4010 \\ 
        7 & 4010 inner loop Q-axis current reference    & \cellcolor{gray!20}2619   & \cellcolor{gray!20}2619 & 4010  \\ 
        8 &  4010 inner loop D-axis current reference   &  \cellcolor{gray!40}1331   &   \cellcolor{gray!20}2619 & 4010 \\ 
        9 & 4010 PLL output angle \color{black}  & 4010   &  4010 & 4010 \\ 
        \bottomrule
    \end{tabularx}
\end{table*}
\color{black}

 In particular, we look at Case $8$ in Table \ref{tab:case_HVDC}, where the FO is injected to the inner loop D-axis current reference of bus $4010$. The $V_{ccp}$ magnitude and angle trajectories are shown in Fig. \ref{fig:HVDC_4010_s1_ang} (a) and (b). 
In addition, the single-sided spectrum from FFT and the associated peak detection observation are also shown in Fig. \ref{fig:HVDC_4010_s1_ang} (c) and (d). 

\textbf{\begin{figure}[!th]
    \centering
    \includegraphics[width=3.2 in]{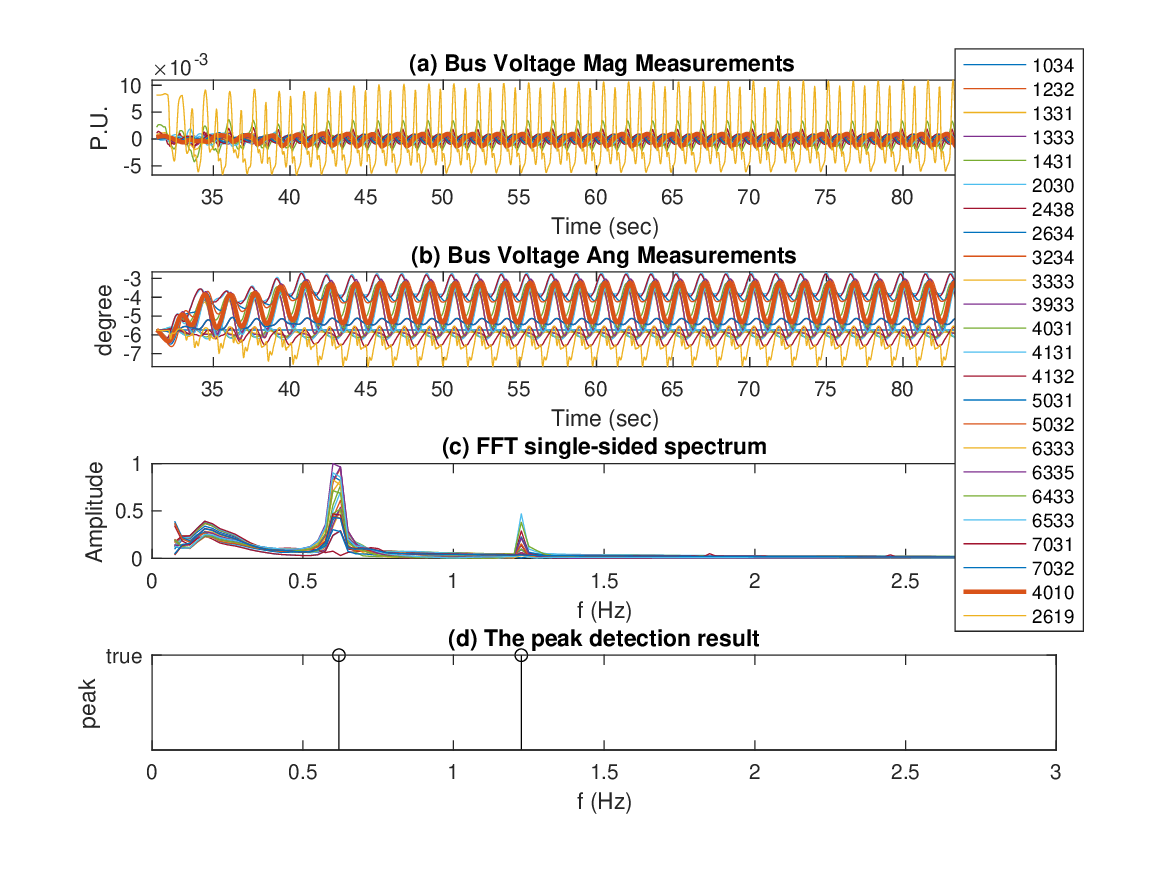}
        \caption{\color{black} (a) $V_{ccp}$ magnitude (b) $V_{ccp}$ angle (c) The single-sided spectrum from FFT (d) Peak frequency detection in the case of forced oscillations originating from the HVDC converter inner loop D-axis current reference at bus $4010$.}
    \label{fig:HVDC_4010_s1_ang}
\end{figure}}
\begin{figure}[!t]
    \centering
    \includegraphics[width=3.4 in]{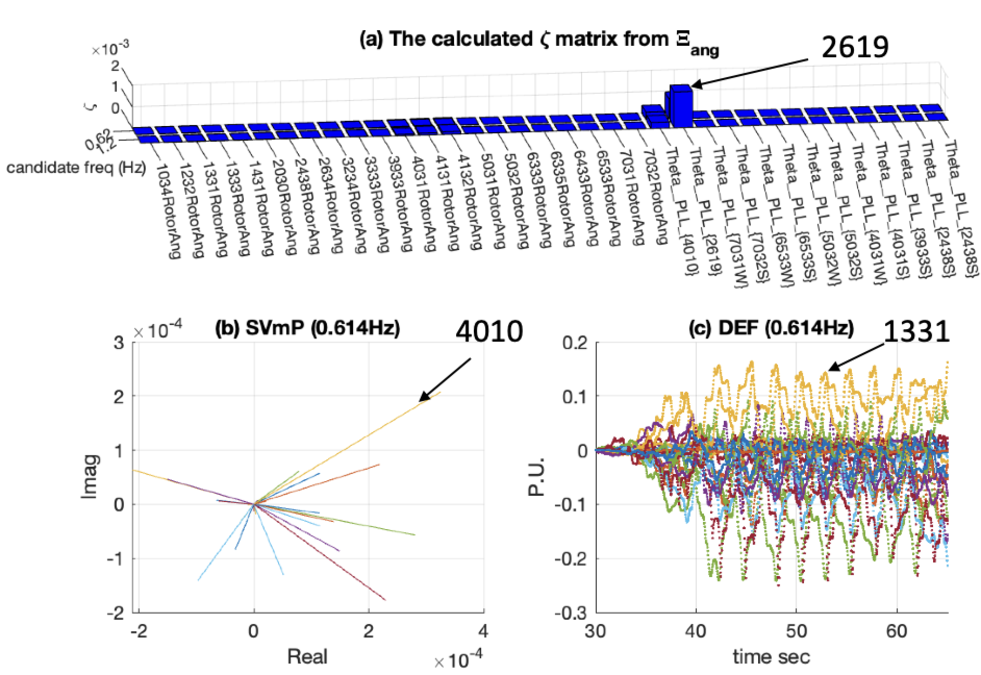}
        \caption{\color{black} (a) Proposed method (b) VMD-CPSD method (c) DEF method in the case of forced oscillations originating from the HVDC converter inner loop D-axis current reference at bus $4010$.}
    \label{fig:HVDC_4010_s1_Xiang}
\end{figure}
Fig. \ref{fig:HVDC_4010_s1_Xiang} illustrates the results of the proposed method in comparison with other methods for this case. The VMD-CPSD method successfully identifies the true source of the oscillation, the sending end bus 4010 of the HVDC, as shown in Fig. \ref{fig:HVDC_4010_s1_Xiang} (b). However, while it correctly narrows the source down to the HVDC system, the proposed method mistakenly identifies the receiving end bus 2619 as the source. In contrast, the DEF method fails to identify the HVDC as the source at all, as shown in Fig. \ref{fig:HVDC_4010_s1_Xiang} (a) and (c).

In general, the results in Table \ref{tab:case_HVDC}  demonstrate that the proposed method tends to misidentify the receiving end bus $2619$ as the FO source rather than the sending end bus $4010$. 
It appears that the $V_{ccp}$ at the sending end 
is less sensitive compared to the $V_{ccp}$ at the receiving end when FO stem from HVDC converter control.  
Further research is required to better understand HVDC dynamics and improve the proposed method for accurately identifying oscillations within HVDC systems. 
\color{black}

\color{blue}

\color{black}

\color{black}
\subsection{Validation under Non-Stationary Forced Oscillations}
A non-stationary FO event caused by renewable generation has an impact on the electromechanical modes of the system \cite{nonStatSurinkaew}.
To test the performance of the proposed method when the forcing frequency and magnitude change over time, simulations are conducted with FO \color{black} of multiple frequency components \color{black} injected into the outer loop of the active power control of renewable generation. 
The oscillations are \color{black}injected every 5 seconds at a single frequency selected among the resonant frequencies \color{black} of $0.379$ Hz, $0.614$ Hz, and $1.27$ Hz and with peak strengths of $1.2\%$, $2.22\%$, and $5.22\%$ of the corresponding active power reference. \color{black}  Consequently, within the 40-second window used by SINDy,
the system experiences FO with mixed frequencies and magnitudes. \color{black}

\begin{figure}[h]
    \centering
    \includegraphics[width=3.4 in]{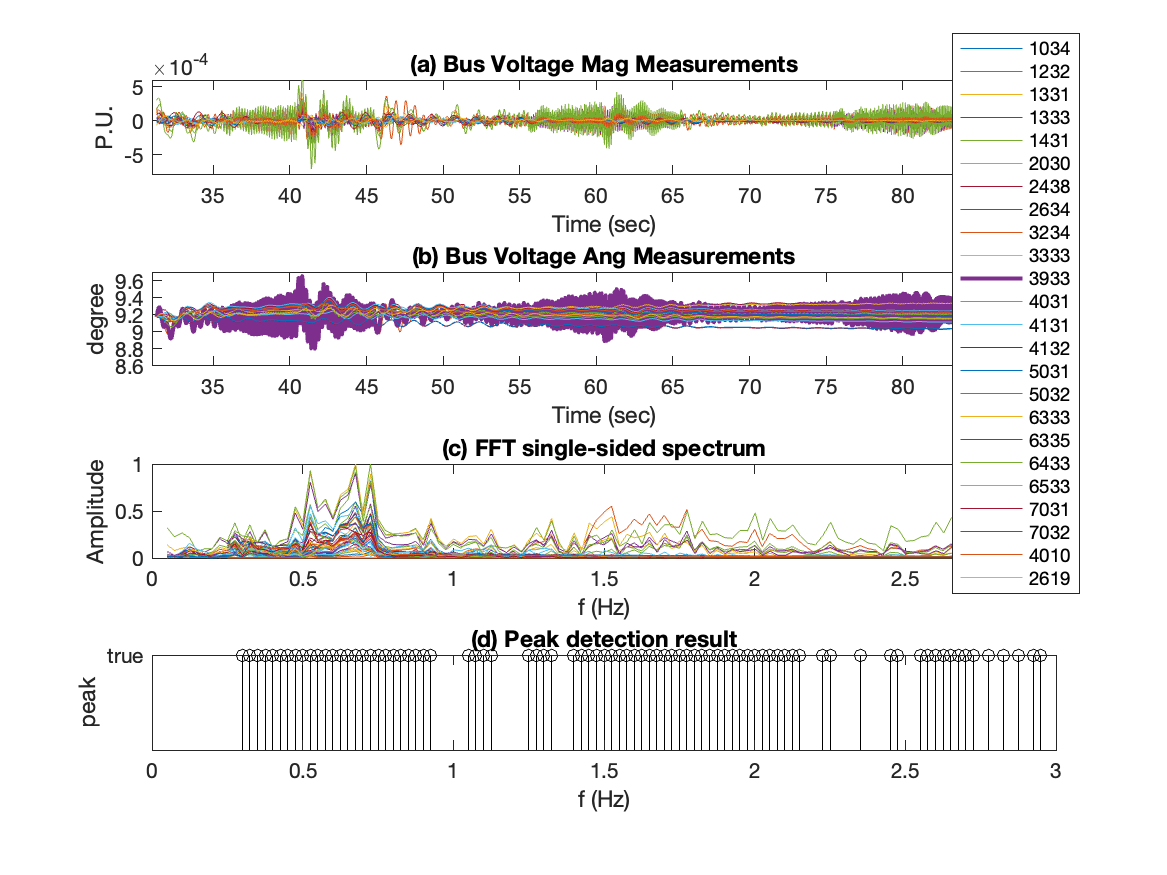}
        \caption{{\color{black} (a) $V_{ccp}$ magnitude (b) $V_{ccp}$ angle (c) The single-sided spectrum from FFT (d) Peak frequency detection in the case of non-stationary forced oscillations originating from solar generation at Bus $3933$.}}
    \label{fig:nonStatFO_3933s_p_ang}
\end{figure}
\begin{figure}[h]
    \centering
    \includegraphics[width=3.2 in]{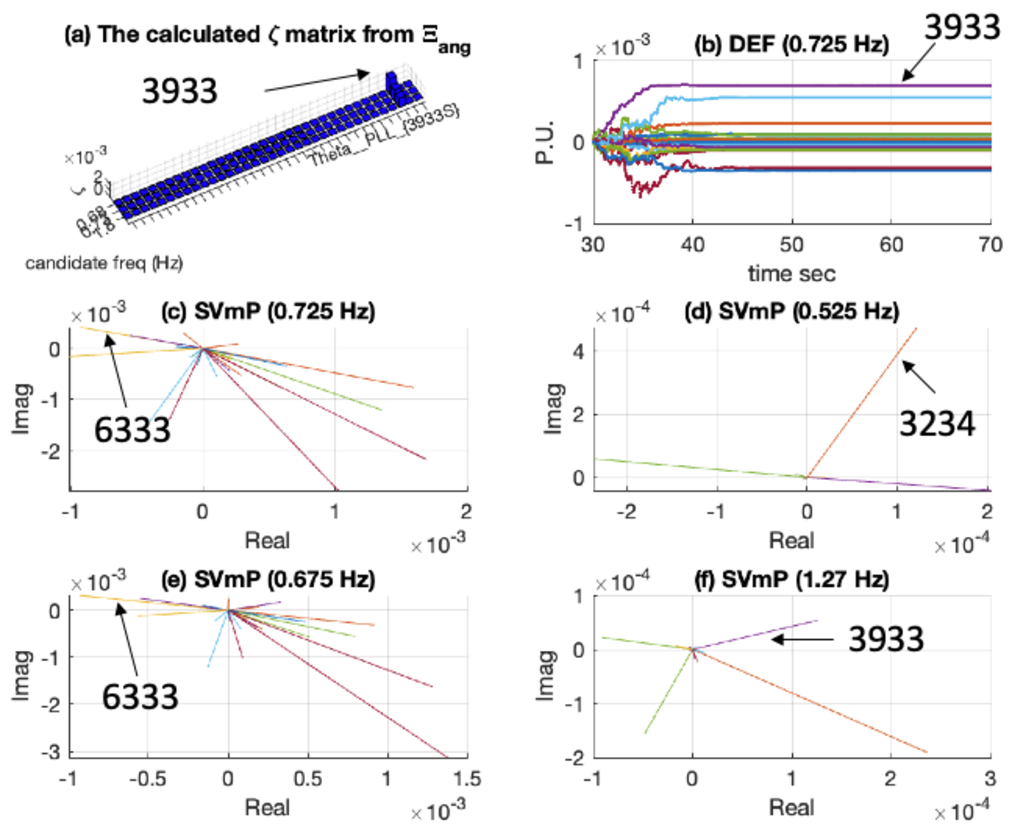}
        \caption{\color{black} (a) Proposed method (b) DEF method (c)-(f) VMD-CPSD method in the case of non-stationary forced oscillations originating from solar generation at Bus $3933$.}
    \label{fig:nonStatFO_3933s_p_VMD}
\end{figure}

The $V_{ccp}$ angle and magnitude trajectories in the case where non-stationary FO originates from solar generation at bus $3933$ are presented in Fig.\,\ref{fig:nonStatFO_3933s_p_ang} (a) and (b). 
The corresponding single-sided spectrum from FFT and peak detection are shown in Fig.\,\ref{fig:nonStatFO_3933s_p_ang} (c) and (d). From Fig. \ref{fig:nonStatFO_3933s_p_ang} (c), 
we select three of the highest amplitude frequencies, namely, $0.725$ Hz, $0.525$ Hz, and $0.675$ Hz, which however, are different from the true FO frequencies. Fig. \ref{fig:nonStatFO_3933s_p_VMD} (a) and (b) illustrate that the proposed method and the DEF method can still correctly identify that there is FO sourced at bus $3933$.  
In contrast, the VMD-CPSD method 
fails to identify the true source when using the incorrect frequencies $0.725$ Hz, $0.525$ Hz, and $0.675$ Hz, since the corresponding highest imaginary parts do not belong to bus $3933$, as shown in Fig. \ref{fig:nonStatFO_3933s_p_VMD} (c) to (e). 
On the other hand, if the correct FO frequencies can be identified, Fig. \ref{fig:nonStatFO_3933s_p_VMD} (f) validates that the VMD-CPSD method successfully locates the source when one of the true oscillation frequencies $1.27$ Hz is used. 

To further study the effect of non-stationary FO of varying frequencies injected in different locations, Table \ref{tab:case_nonStatFO} provides the source locating results when using two of the true oscillation frequencies, specifically $1.27$ Hz and $0.379$ Hz. The simulation results suggest that the proposed SINDy method can accurately capture the true FO source whereas the other two methods may fail or render inconsistent results even when true FO frequencies are applied. 

\begin{figure}[!th]
    \centering
    \includegraphics[width=3.4 in]{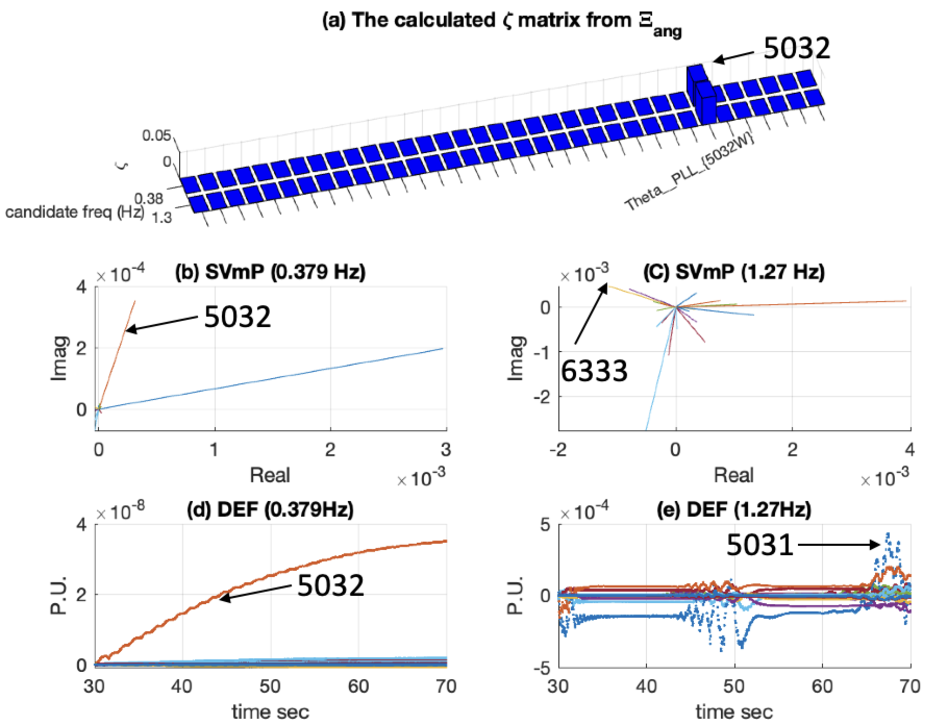}
        \caption{\color{black} (a) Proposed method (b)-(c) VMD-CPSD method (d)-(e) DEF method in the case of non-stationary forced oscillations originating from solar generation at Bus $5032$.}
    \label{fig:nonStat5032}
\end{figure}

For example, we consider Case $5$ in Table \ref{tab:case_nonStatFO} where the non-stationary FO is added to the active power control reference of the solar generation at Bus 5032. From Fig. \ref{fig:nonStat5032}, it can be observed that when applying the DEF and VMD-CPSD methods using the true FO frequency of $1.27$ Hz, the results can turn out incorrect. 
This situation highlights the challenges for the DEF and VMD-CPSD methods when varying frequencies are present, as they are designed to accommodate only one FO frequency at a time. In those situations, the proposed SINDy may complement the VMD-CPSD and DEF methods in FO locating. 
\color{black} 
\small
\begin{table}[!ht]
    \caption{\color{black}Non-stationary forced oscillation cases originating from solar/wind generation}
    \label{tab:case_nonStatFO}
    \centering
    \begin{tabular}{c c c c c c c}
    \toprule
    \multicolumn{2}{|c|}{\makecell{Simulation\\setup}} & \multicolumn{2}{c|}{DEF}& \multicolumn{1}{c|}{SINDy}& \multicolumn{2}{c|}{VMD-CPSD} \\ 
    \midrule
        {\#} & \makecell{Injection\\bus}   &  \makecell{ Est at\\0.379Hz }    & \makecell{ Est at\\1.27Hz }    &  \makecell{ Est with\\0.379Hz,\\
        1.27Hz} & \makecell{ Est at\\0.379Hz }  & \makecell{ Est at\\1.27z }  \\ \midrule
        {1} &2438 solar    & 2438 & \cellcolor{gray!40}5031 & 2438   &  2438   & 2438  \\ 
        {2} &3933 solar     & 3933& 3933& 3933   &  3933   & 3933  \\ 
        {3} &4031 solar    &4031 &4031  & 4031   &  4031   & 4031  \\ 
        {4} &4031 wind    & 4031 & \cellcolor{gray!40}5031 & 4031   &  4031   & 4031  \\ 
        {5} &5032 solar    & 5032 & \cellcolor{gray!20}5031 & 5032   &  5032   & \cellcolor{gray!40}6333  \\ 
        {6} &5032 wind     & 5032 & \cellcolor{gray!20}5031 & 5032   &  5032   & \cellcolor{gray!40}6333  \\ 
        {7} &6533 solar    & 6533 & \cellcolor{gray!20}6333 & 6533   &  \cellcolor{gray!20}6333   & \cellcolor{gray!40}4031  \\ 
        {8} &6533 wind    & 6533 & 6533 & 6533   &  6533   & \cellcolor{gray!40}4031  \\ 
        {9} &7032 solar  & 7032 & 7032 & 7032   &  7032   & 7032  \\   
        \bottomrule
    \end{tabular}
\end{table}
\normalsize

\color{black}
\subsection{Locating Forced Oscillations from Synchronous Generators}

\begin{figure}[h]
    \centering
    \includegraphics[width=3.5 in]{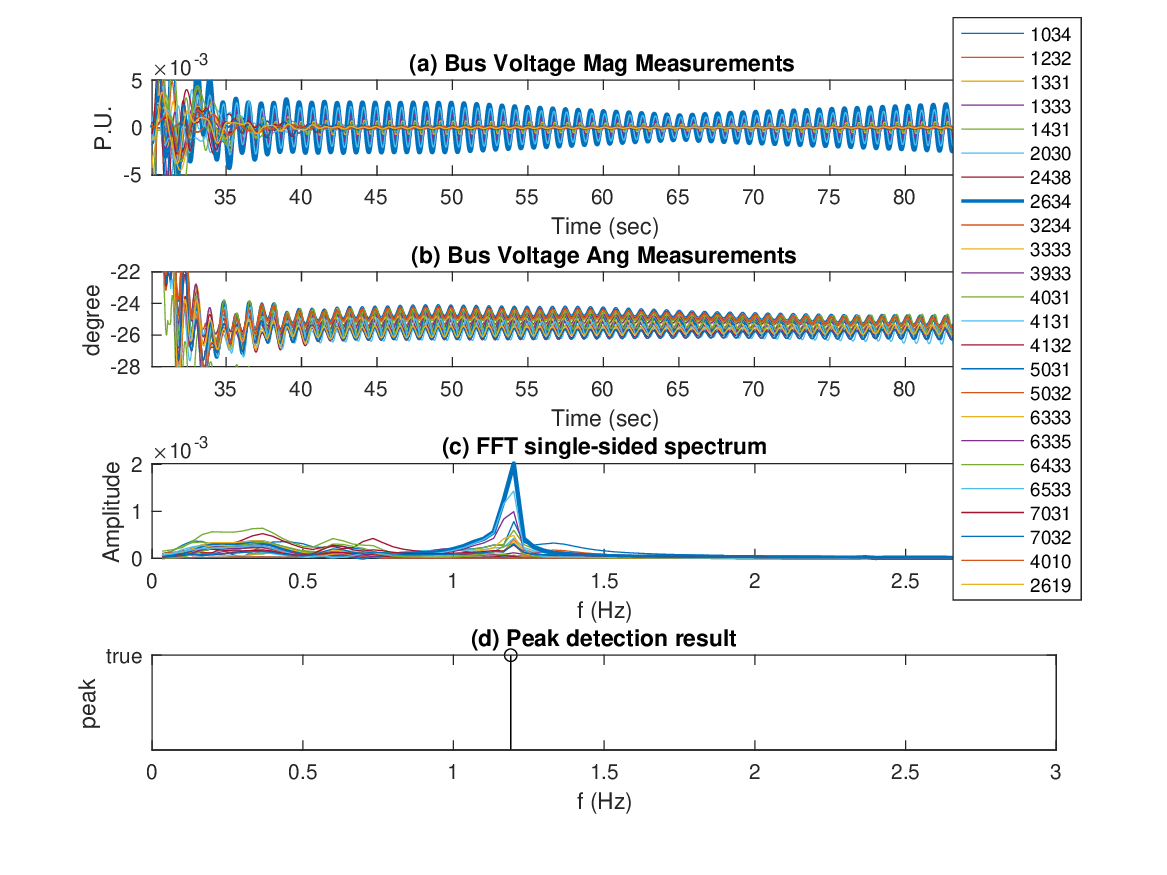}
        \caption{(a) $V_{ccp}$ magnitude (b) $V_{ccp}$ angle (c) The single-sided spectrum from FFT (d) Peak frequency detection in the case of forced oscillations originating from synchronous generation at bus $2634$.}
    \label{fig:StatFO_2634_ang}
\end{figure}

Previous sections have mostly focused on FO sourced from IBRs. To further validate the performance of the proposed method in systems hosting both conventional generators and IBRs, in this section we present the results for the case where the FO source is a conventional synchronous generator. 
A simulation of $90$ seconds is carried out \color{black}and all synchronous generators are modeled using a sixth-order model—the ``GENROU" model in TSAT—which includes rotor angle, rotor speed, and transient and sub-transient dynamics. Detailed excitation systems and turbine governors are also incorporated. 
\color{black}
A forced oscillation of $1.2$ Hz is injected into the active power reference of the synchronous generator on bus $2634$ at $30$ seconds. The FO strength is $5\%$ of the corresponding active power reference value. 
All real power generations in the system experience 50db random noise ($1\%$ of the forced oscillation) as processing noise.
The time window for the collection of the measurements is selected to be $40$ seconds after the injection with a reporting rate of $60$ frames per second.
The voltage and angle trajectories of the AC bus that is immediately connected to the monitored synchronous generator rotor and the results of the FFT and peak frequency detection are presented in Fig. \ref{fig:StatFO_2634_ang} (a)--(d). 

\begin{figure}[h]
    \centering
    \includegraphics[width=3.3 in]{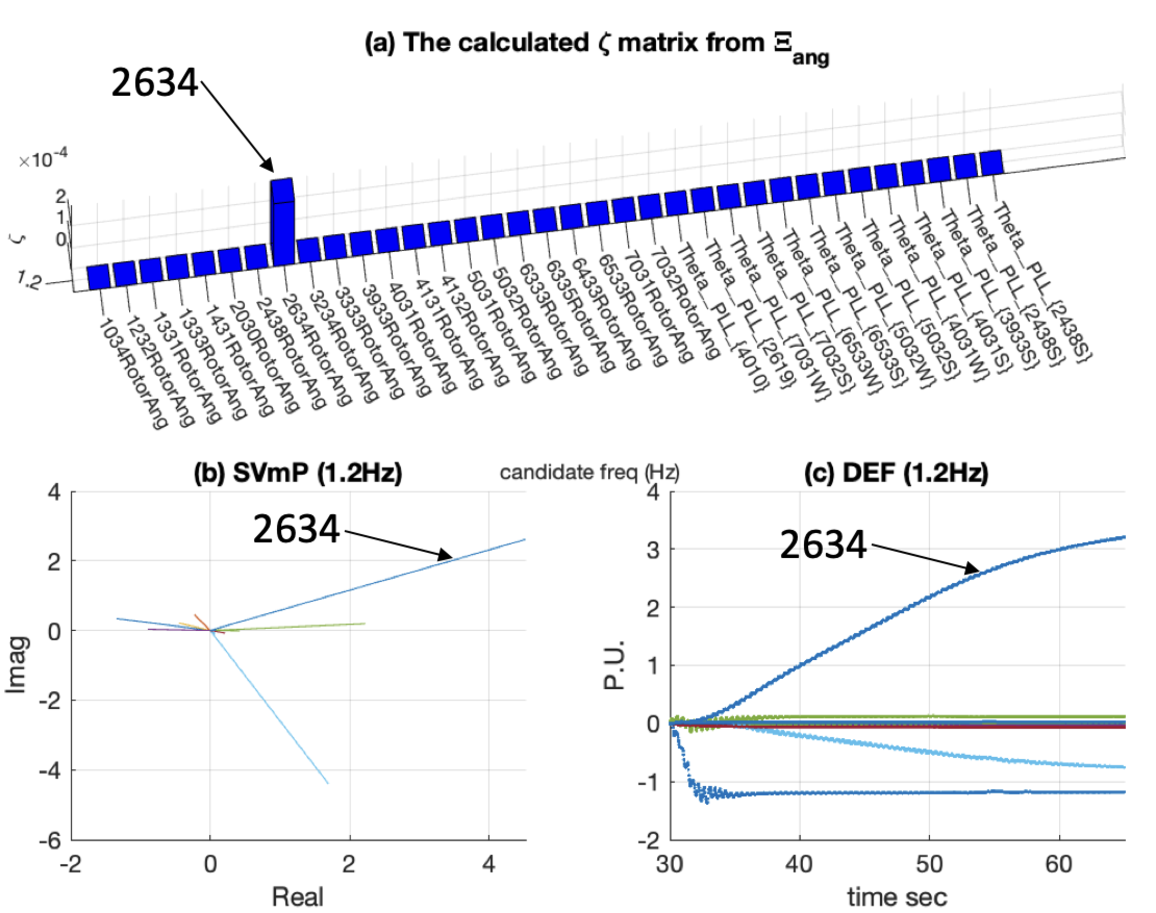}
        \caption{(a) Proposed method (b) VMD-CPSD method (c) DEF method in the case of forced oscillations originating from synchronous generation at bus $2634$.}
    \label{fig:StatFO_2634_Result}
\end{figure}

Fig. \ref{fig:StatFO_2634_Result} (a) shows that the proposed algorithm determines that the source is located at bus $2634$ as the highest peak appears on the elements that belong to the rotor angle of the synchronous generator connected to the bus $2634$ and the frequency of $1.2$ Hz. 
This result agrees with the VMD-CPSD and DEF algorithm as shown in Fig. \ref{fig:StatFO_2634_Result} (b) and (c). 

\color{black}More results can be found in Table \ref{tab:SINDyComparison1}, which shows the source localization results for FO injected at various generators through both exciters and governors.  
These results show that the proposed algorithm can successfully locate the true source location when the FO originates from conventional synchronous generators. 
Further supporting results can also be found in \cite{cai2022}.
\color{black}

\begin{table}[!ht]
\color{black}
    \caption{Forced oscillation cases with source from the synchronous generator}
    \label{tab:SINDyComparison1}
    \centering
     \begin{tabular}{c c c c c c c}
        \toprule
        \multicolumn{3}{|c|}{\makecell{Simulation\\setup}} & \multicolumn{1}{c|}{DEF}& \multicolumn{1}{c|}{SINDy}& \multicolumn{1}{c|}{VMD-CPSD} \\ 
        \midrule
         {\#}  &  FO Description  &  Inj bus  &  Est bus  & Est bus  &  Est bus \\ \midrule
        1 & Governor at 0.82 Hz &  1431  & 1431   &  1431   & 1431  \\ 
        2 & Governor at 1.19 Hz &  2634  & 2634   &  2634   & 2634 \\
        3 & Governor at 1.27 Hz &  7031  & 7031  & 7031 & 7031 \\ 
        4 & Exciter at 0.379 Hz &  1331  & 1331   &  1331   & 1333  \\
        5 & Exciter at 0.614 Hz &  6333  & 6333   &  6333 & 6333 \\ 
        \bottomrule
    \end{tabular}
\end{table}

\section{Conclusions}
\label{section:6}
\color{black}
This paper proposes a purely data-driven algorithm for locating Forced Oscillation (FO) sources in the presence of \color{black} grid-following (GFL) \color{black} Inverter Based Resources (IBRs). 
Leveraging on Sparse Identification of Nonlinear Dynamics (SINDy), the proposed algorithm can build a linear mapping between measurements and a library filled with suitable FO source candidates including the classic generator model as well as a suitable VSC PLL dynamic model. 
Simulations of injecting FO in the WECC 240-bus system verify the effectiveness of the proposed algorithm in successfully locating the true FO sources under resonance conditions and non-stationary FO of various frequencies. Interestingly, it is validated that the proposed approach outperforms existing methods in the case of FO consisting of multiple frequencies. 
Future research will focus on further analyzing and pinpointing the location of simultaneous FO sources; how the proposed method can be extended to integrate grid-forming IBRs and their impact on system dynamics. \color{black}Another promising direction is combining SINDy with physics-informed methods like DEF to improve robustness, where DEF can help regularize SINDy and mitigate its sensitivity to noise.
\color{black}


\bibliographystyle{IEEEtran}
\typeout{}
\bibliography{ref.bib}

\begin{IEEEbiography}[{\includegraphics[width=1in,height=1.25in,clip,keepaspectratio]{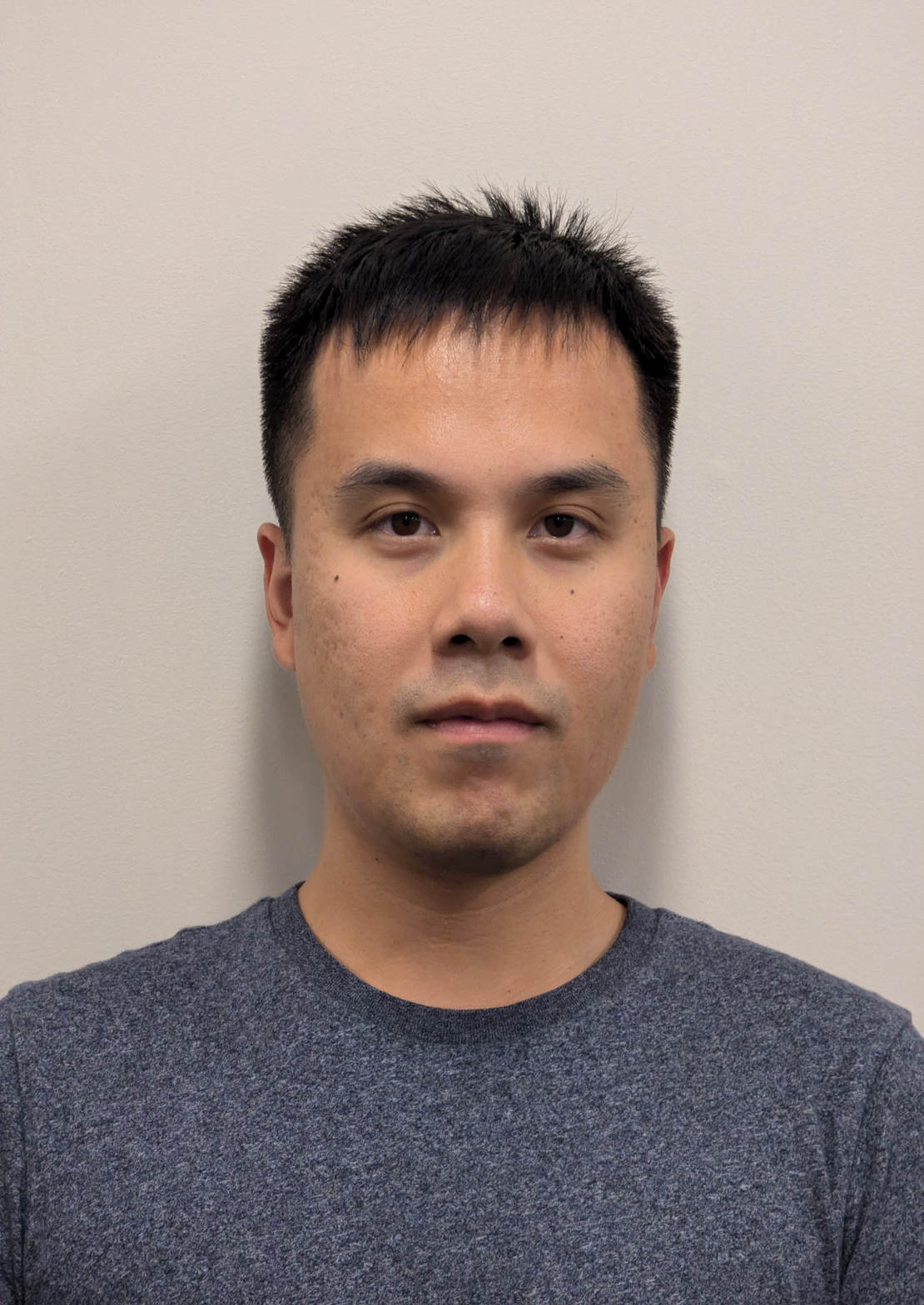}}]{Yaojie Cai}
received the MSc degrees in electrical and computer engineering from the University of Manitoba, Winnipeg, MB, Canada. He is currently working toward the Ph.D. degree in electrical and computer engineering with McGill University, QC, Canada. His research interests include data analytics, power system stability and control.
\end{IEEEbiography}

\begin{IEEEbiography}[{\includegraphics[width=1in,height=1.25in,clip,keepaspectratio]{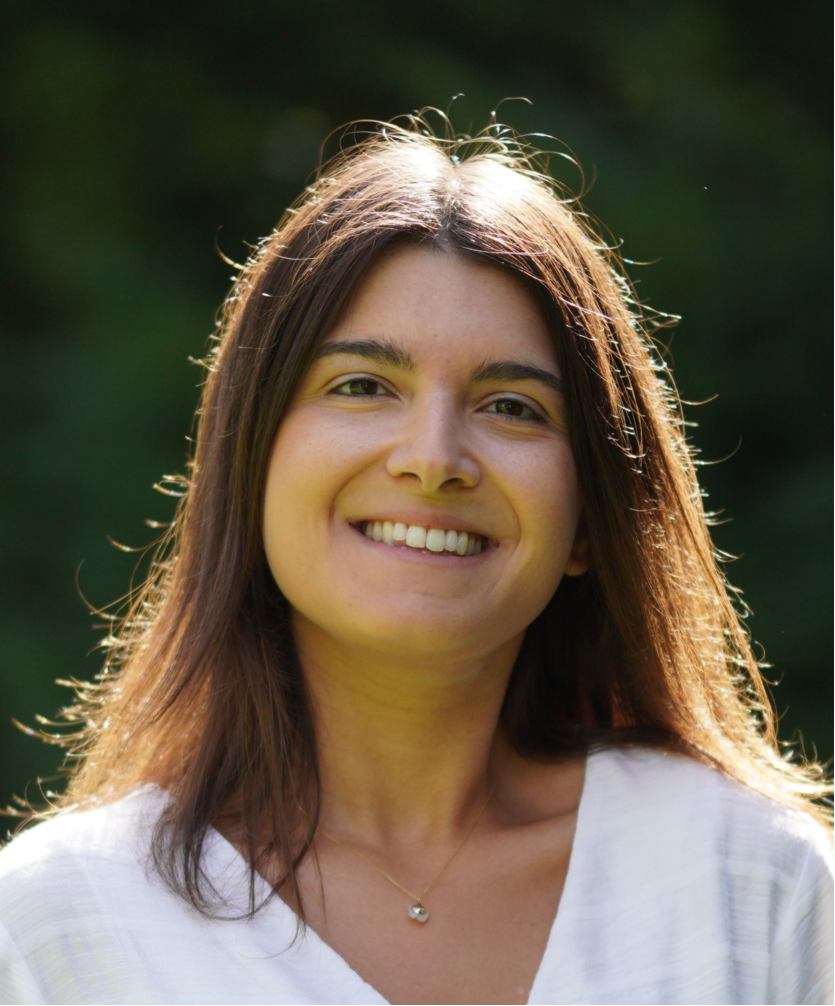}}]{Georgia Pierrou }
(Member, IEEE) received the Diploma in electrical and computer engineering from the National Technical University of Athens, Athens, Greece in 2017, and the Ph.D. degree from McGill University, Montreal, QC, Canada, in 2021. She is currently a Postdoctoral Researcher with the Power Systems Laboratory, ETH Zurich, Zurich, Switzerland. Her research interests include dynamic analysis, optimization, and control of electric power systems.
\end{IEEEbiography}

\begin{IEEEbiography}[{\includegraphics[width=1in,height=1.25in,clip,keepaspectratio]{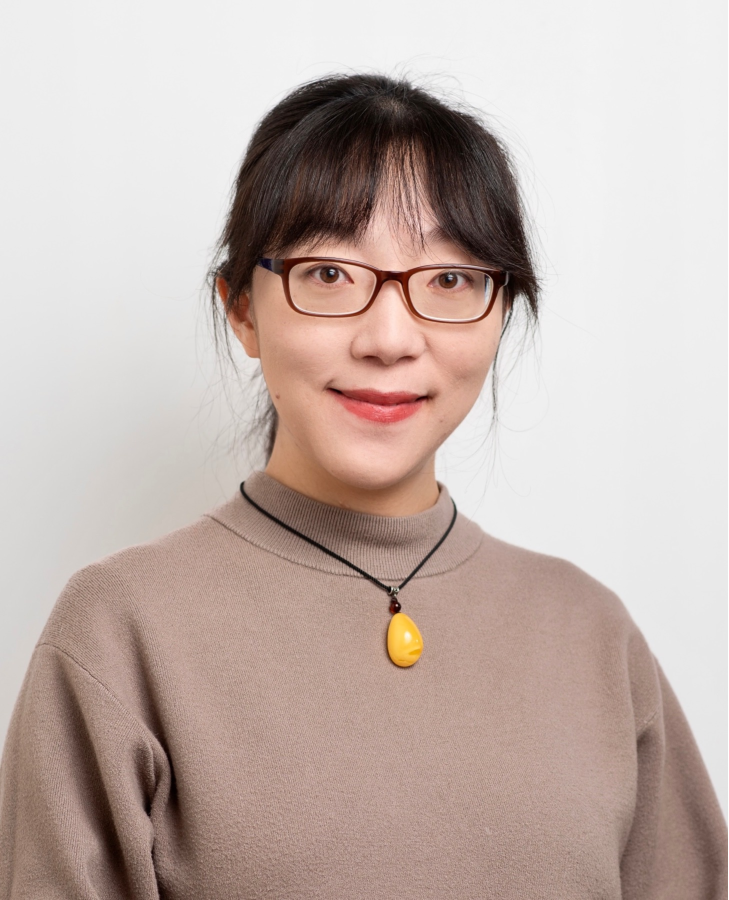}}]{Xiaozhe Wang}
(Senior Member, IEEE) received the B.S. degree in information science and electronic engineering from Zhejiang University, Zhejiang, China, in 2010, and the Ph.D. degree from the School of Electrical and Computer Engineering, Cornell University, Ithaca, NY, USA, in 2015. She is currently an Associate Professor and a Canada Research Chair with the Department of Electrical and Computer Engineering, McGill University, Montreal, QC, Canada. Her research interests include data-driven power system stability monitoring and control, uncertainty quantification in power system security, stability and resilience, and cybersecurity in power systems.
\end{IEEEbiography}

\begin{IEEEbiography}[{\includegraphics[width=1in,height=1.25in,clip,keepaspectratio]{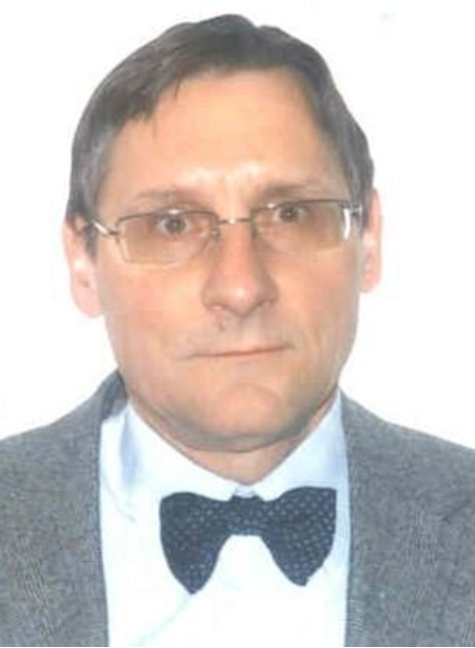}}]{Géza Joós}
(Life Fellow, IEEE) graduated from McGill University, Montreal, Canada, with an M.Eng. and Ph.D. in Electrical Engineering. He is a Professor in the Department of Electrical and Computer Engineering Department at McGill University (since 2001). He holds a Canada Research Chair in Powering Information Technologies (since 2004). His research interests are in distributed energy resources, including renewable energy resources, advanced distribution systems and microgrids. He was previously with ABB, the Université du Québec and Concordia University (Montreal, Canada). He is active in IEEE Standards Association working groups on distributed energy resources and microgrids. He is a Fellow of CIGRE, and the Canadian Academy of Engineering.
\end{IEEEbiography}

    \end{document}